\documentclass[twocolumn,longbib]{aastex701}

\usepackage{makecell,CJK}

\newcommand{\jwst}{JWST}

\newcommand{\rev}[1]{{{\color{black}{}#1}}}

\newcommand{\auburn}{Department of Physics, Auburn University, Edmund C.\ Leach Science Center, Auburn, 36849, AL, USA}
\newcommand{\esaneocc}{ESA PDO NEO Coordination Centre, Largo Galileo Galilei, 1, I-00044 Frascati (RM), Italy}
\newcommand{\jpl}{Jet Propulsion Laboratory, California Institute of Technology, 4800 Oak Grove Dr., Pasadena, CA 91109, USA}
\newcommand{\lowell}{Lowell Observatory, 1400 W.\ Mars Hill Rd., Flagstaff, AZ 86001, USA}
\newcommand{\planetaryscienceinst}{Planetary Science Institute, 1700 East Fort Lowell Rd., Suite 106, Tucson, AZ 85719, USA}
\newcommand{\umd}{Department of Astronomy, University of Maryland, 4296 Stadium Dr., College Park, MD 20742, USA}
\newcommand{\villanova}{Department of Astrophysics and Planetary Science, Villanova University, Villanova, PA, 19085, USA}
\newcommand{\carnegie}{Earth and Planets Laboratory, Carnegie Institution for Science, 5241 Broad Branch Road NW, Washington, DC 20015, USA}
\newcommand{\uwdirac}{Department of Astronomy \& the DiRAC Institute, University of Washington, 3910 15th Ave NE, Seattle, WA 98195, USA}
\newcommand{\lincc}{LSST Interdisciplinary Network for Collaboration and Computing, 933 N.\ Cherry Avenue, Tucson, AZ 85721, USA}
\newcommand{\nau}{Department of Astronomy \& Planetary Science, Northern Arizona University, P.O.\ Box 6010, Flagstaff, AZ 86011, USA}
\newcommand{\rawdata}{Raw Data Speaks Initiative, USA}
\newcommand{\edinburgh}{Institute for Astronomy, University of Edinburgh, Royal Observatory, Edinburgh EH9 3HJ, UK}

\providecommand\aap{Astron.\ Astrophys.} 
\providecommand\apjl{Astrophys.\ J.} 
\providecommand\apjs{Astrophys.\ J.\ Suppl.} 

\received{December 8, 2025}
\accepted{April 9, 2026}
\submitjournal{PSJ}

\shorttitle{\jwst{} and Ground-Based Observations of 457P}
\shortauthors{Noonan et al.}

\hypersetup{pdfauthor={Name}}
\begin{document}
\begin{CJK*}{UTF8}{gbsn}

\title{Emerging Diversity Among the Main-Belt Comets:\\Insights from \jwst{} and Ground-Based Observations of 457P/Lemmon-PANSTARRS}

\correspondingauthor{John Noonan}
\email{noonan@auburn.edu}

\author[0000-0003-2152-6987]{John W.\ Noonan}
\affiliation{\auburn}
\email{noonan@auburn.edu}

\author[0000-0001-7225-9271]{Henry H.\ Hsieh}
\affiliation{\planetaryscienceinst}
\email{hhsieh@psi.edu}

\author[0000-0002-6702-7676]{Michael S.\ P.\ Kelley}
\affiliation{\umd}
\email{msk@astro.umd.edu}

\author[0000-0002-2668-7248]{Dennis Bodewits}
\affiliation{\auburn}
\email{dzb0059@auburn.edu}

\author[0000-0002-5736-1857]{Jana Pittichov\'a}
\affiliation{\jpl}
\email{jana.chesley@jpl.nasa.gov}

\author[0000-0002-1506-4248]{Audrey Thirouin}
\affiliation{\lowell}
\email{thirouin@lowell.edu}

\author[0000-0001-7895-8209]{Marco Micheli}
\affiliation{\esaneocc}
\email{marco.micheli@esa.int}

\author[0000-0003-3145-8682]{Scott S.\ Sheppard}
\affiliation{\carnegie}
\email{ssheppard@carnegiescience.edu}

\author[0000-0001-7335-1715]{Colin O.\ Chandler}
\affiliation{\uwdirac}
\affiliation{\lincc}
\affiliation{\nau}
\affiliation{\rawdata}
\email{coc123@uw.edu}

\author[0000-0003-1008-7499]{Theodore Kareta}
\affiliation{\lowell}
\affiliation{\villanova}
\email{theodore.kareta@villanova.edu}

\author[0000-0001-9328-2905]{Colin Snodgrass}
\affiliation{\edinburgh}
\email{csn@roe.ac.uk}

\author[0009-0007-5946-8731]{Richard E. Cannon}
\affiliation{\edinburgh}
\email{richard.cannon@ed.ac.uk}

\author[0000-0002-8137-5132]{Brian P. Murphy}
\affiliation{\edinburgh}
\email{brian.murphy@ed.ac.uk}

\begin{abstract}
We present JWST NIRSpec and NIRCam observations of 457P/Lemmon-PANSTARRS, a main-belt comet that displayed activity around its 2020 perihelion and that was observed to regain activity during its 2024 perihelion by a ground-based observing campaign. The previous successful measurements of water production from two main-belt comets by the JWST NIRSpec instrument confirmed the hypothesis that H$_2$O reservoirs are responsible for activity in dynamically stable main-belt comets.
However, the main-belt comets observed with JWST thus far, 238P/Read and 358P/PANSTARRS, occupy orbits in the outer main-belt, with main-belt comets with smaller semi-major axes not yet sensitively tested for H$_2$O. We find that despite clearly displaying dust activity in both ground-based and JWST imaging over a broad period, there were no corresponding H$_2$O, CO, CO$_2$, or CH$_3$OH emissions within sensitive upper limits; notable given 457P is the first main-belt comet with a semi-major axis within the 5:2 mean-motion resonance with Jupiter. We show that we were sensitive to production rates of gas predicted by the dust/gas ratios of 238P and 358P, and hypothesize that 457P may be more depleted than its companions; Q$_{H2O}$ must be less than 2$\times10^{24}$ molecules sec$^{-1}$, or 0.035 kg sec$^{-1}$. Further surveying of main-belt comets across the parameter space of semi-major axis and eccentricity will shed light on whether 457P represents an edge member of a spectrum or a distinct subclass of main-belt comets.. 
\end{abstract}


\section{Introduction\label{section:intro}}
\setcounter{footnote}{0}

\subsection{Background\label{section:background}}

Main-belt comets (MBCs) occupy a unique position within the solar system's remnant planetesimals, with orbits constraining them to the main-belt of asteroids but repeated displays of near-perihelion activity that make them more akin to comets \citep{hsieh2006_mbcs,snodgrass2017_mbcs, jewitt2024_continuum_comets3}. MBCs are a subset of active asteroids, which the broader category of objects that display comae but have Tisserand parameters $>$3 \citep{jewitt2024_continuum_comets3}, but this periodic reactivation makes them distinct. The source of this activity was hypothesized to be due to water sublimation, as dust production was closely linked to the MBCs' true anomaly, which occurred during multiple apparitions and typically peaks near perihelion, much like a standard comet, and was inconsistent with other activations mechanisms like rotational shedding or disruption. This hypothesis was confirmed by \jwst{} observations of 238P/Read and 358P/PANSTARRS, which detected the fundamental $\nu_1$ and $\nu_3$ vibrational bands of H$_2$O emission using the NIRSpec instrument \citep{kelley2023_jwst238p,hsieh2025_358p}. These detections demonstrated that the water production rates of these MBCs were elusive for other ground- or space-based observatories, including the Hubble Space Telescope (HST), due to both sensitivity and timing constraints. While typically observed cometary water production rates span 10$^{27}$ to 10$^{31}$ molecules s$^{-1}$ \rev{\citep{ahearn1995_ensemblecomets,combi2019_swancomets}}, the values measured for these MBCs were roughly two orders of magnitude lower.

The presence of water ice in dynamically stable main-belt objects, where any permanently exposed ice should have sublimated long ago, challenges previous commonly-held assumptions about the distribution and long-term retention of volatiles in the asteroid belt. Given the inefficiency of implanting Jupiter family comets (JFCs) from the outer solar system onto low-inclination main-belt orbits given the solar system's current dynamical architecture, the more likely hypothesis is that comets like 238P and 358P formed near their current locations with somewhat substantial reservoirs of water ice or were at least delivered to those locations early in the solar system's history \citep{hsieh2016_tisserand}. How close they formed, however, and whether they were implanted from a different reservoir far earlier, is still a key question. 
As posited by \cite{hsieh2006_mbcs}, this reservoir could be preserved over the lifetime of our solar system beneath 1-100 m of regolith. For objects such as 238P and 358P, with radii in the 200-350 m range, this represents a substantial fraction of their overall volume \citep{hsieh2009_238p,hsieh2016_238p,agarwal2018_358pnucleus,hsieh2018_358p}. To assess if our current understanding of MBC activity as the result of weak sublimation, and in particular to determine if dust-rich comae are ubiquitous among the population, it is necessary to continue the search for H$_{2}$O among the known MBCs.

To further explore the properties of main-belt comets, we obtained \jwst{}  observations of the MBC 457P/PANSTARRS. NIRCam and NIRSpec observations were taken on UT 2024 September 20, thirty days after perihelion, optimized to capture peak activity. In this manuscript, we will discuss the JWST observations, supporting ground campaign, dust and volatile analysis, and implications of our findings. 

\setlength{\tabcolsep}{5pt}
\setlength{\extrarowheight}{0em}
\begin{table*}[htb!]
\caption{\jwst{} 457P Observations$^a$}
\centering
\smallskip
\footnotesize
\begin{tabular}{clccccccccccc}
\hline\hline
\multicolumn{1}{c}{Target}
 & \multicolumn{1}{c}{UT Date}
 & \multicolumn{1}{c}{UT Time}
 & \multicolumn{1}{c}{Instrument}
 & \multicolumn{1}{c}{$\nu$$^b$}
 & \multicolumn{1}{c}{$r_h$$^c$}
 & \multicolumn{1}{c}{$\Delta_{\oplus}$$^d$}
 & \multicolumn{1}{c}{$\Delta_{obs}$$^e$}
 & \multicolumn{1}{c}{$\alpha_{\oplus}$$^f$}
 & \multicolumn{1}{c}{$\alpha_{\rm obs}$$^g$}
 & \multicolumn{1}{c}{PA$_{-\odot}$$^h$}
 & \multicolumn{1}{c}{PA$_{-v}$$^i$}
 & \multicolumn{1}{c}{$\Delta t_q$$^j$}
 \\
\hline
457P & 2024 Sep 20 & 17:42:17 - 18:26:19 & NIRCam  &  9.2 & 2.335 & 2.176 & 2.175 & 25.4 & 25.6 &  92.1 & 272.4 & $+$30 \\
457P & 2024 Sep 20 & 14:56:47 - 20:07:17 & NIRSpec &  9.2 & 2.335 & 2.175 & 2.175 & 25.4 & 25.6 &  92.1 & 272.4 & $+$30 \\
\hline
\hline
\multicolumn{13}{l}{$^a$ Observing geometry parameters from JPL Horizons (Orbit solution \#12) \citep{giorgini1996_horizons}} \\
\multicolumn{13}{l}{$^b$ True anomaly, in degrees} \\
\multicolumn{13}{l}{$^c$ Heliocentric distance, in au} \\
\multicolumn{13}{l}{$^d$ Geocentric distance, in au} \\
\multicolumn{13}{l}{$^e$ \jwst{}-centric distance, in au} \\
\multicolumn{13}{l}{$^f$ Solar phase angle (Sun-target-Earth), in degrees} \\
\multicolumn{13}{l}{$^g$ Solar phase angle (Sun-target-\jwst{}), in degrees} \\
\multicolumn{13}{l}{$^h$ Position angle of the anti-Solar vector as projected on the sky, in degrees East of North} \\
\multicolumn{13}{l}{$^i$ Position angle of the negative heliocentric velocity vector as projected on the sky, in degrees East of North} \\
\multicolumn{13}{l}{$^j$ Time relative to perihelion (positive values indicating time after perihelion), in days.} \\
\end{tabular}
\label{table:jwst_457p_observations}
\end{table*}

\subsection{457P/Lemmon-PANSTARRS\label{section:457P_background}}

457P/Lemmon-PANSTARRS was discovered to be active as P/2020 O1 by the 1.8~m Pan-STARRS1 survey telescope on Haleakala \citep{weryk2020_p2020o1} on UT 2020 July 20, after being independently discovered and reported as an apparently asteroidal object by the 1.5~m Mt.\ Lemmon Survey telescope on UT 2020 July 19.  \rev{Pre-discovery observations taken from the 8.2-m Subaru telescope showed that there was also a narrow tail visible in images obtained 3 July 2016 ($\nu$=48$^{\circ}$), but not on 1 August 2016 ($\nu$=56$^{\circ}$), indicating at least three apparitions with activity near perihelion \citep{ly2023_457p_cbet}.}
With a semimajor axis of $a=2.646$~au, perihelion distance of $q=2.329$~au, eccentricity of $e=0.120$, inclination of $i=5.222^{\circ}$, and a Tisserand parameter with respect to Jupiter of $T_J=3.376$, it is unambiguously a main-belt asteroid from a dynamical perspective, but with its observed activity, it is also one of the $\sim$50 currently known active asteroids \citep[see][]{jewitt2024_continuum_comets3}.
HST observations of the object during its 2020 active apparition showed long-duration mass loss (i.e., lasting $\sim$180 days), consistent with sublimation-driven activity \citep{kim2022_p2020o1}, meaning that it therefore is also considered a MBC.

In the work by \citet{kim2022_p2020o1}, the nucleus of 457P was determined to have an absolute magnitude of $H_V=19.25\pm0.13$ (assuming a IAU phase function slope parameter of $G=0.15$, consistent with C-type asteroids) and an estimated effective circular radius of $r_n=420$~m (assuming a geometric albedo of $p_V=0.05$), and was also tentatively identified as a fast rotator with a double-peaked period of $P_{rot}\sim1.7$~hr.  Dust modeling indicated that the activity was consistent with dust grains ranging in radii from $\sim$0.8\,$-$\,1.14~mm with terminal ejection velocities of $v_r\sim0-0.3$~m~s$^{-1}$.

\section{Observations\label{section:observations}}

\jwst{} \citep{gardner2023_jwst} observations of 457P were obtained by NIRCam
\citep{rieke2023_nircam}
 and by NIRSpec 
\citep{jakobsen2022_nirspec,boker2023_nirspec}
on UT 2024 September 20 as part of \jwst{} General Observer (GO) program GO 5551\footnote{\url{https://www.stsci.edu/jwst/science-execution/program-information?id=5551}} in Cycle 3.  Observational circumstances of these observations are listed in Table~\ref{table:jwst_457p_observations}.

\setlength{\tabcolsep}{10pt}
\setlength{\extrarowheight}{0em}
\begin{table*}[htb!]
\caption{Ground-Based Observing Instrumentation Characteristics}
\centering
\smallskip
\footnotesize
\begin{tabular}{lcccc}
\hline\hline
\multicolumn{1}{c}{Telescope$^a$}
 & \multicolumn{1}{c}{Instrument}
 & \multicolumn{1}{c}{FOV$^b$}
 & \multicolumn{1}{c}{Pixel Scale}
 & \multicolumn{1}{c}{Binning}
 \\[2pt]
\hline
Magellan & IMACS   & $15\farcm4\times15\farcm4$ & $0\farcs20$  & $1\times1$ \\
Gemini-N & GMOS-N  & $5\farcm5\times5\farcm5$   & $0\farcs16$  & $2\times2$ \\
Gemini-S & GMOS-S  & $5\farcm5\times5\farcm5$   & $0\farcs16$  & $2\times2$ \\
NTT      & EFOSC2  & $3\farcm9\times3\farcm9$   & $0\farcs24$  & $1\times1$ \\
Palomar  & WaSP    & $18\farcm4\times18\farcm5$ & $0\farcs175$ & $1\times1$ \\
LDT      & LMI    & $12\farcm3\times12\farcm3$ & $0\farcs24$  & $2\times2$ \\
\hline
\hline
\multicolumn{5}{l}{$^a$ Magellan: Magellan Baade telescope; Gemini-N: Gemini North} \\
\multicolumn{5}{l}{$~~~$ telescope; Gemini-S: Gemini South telescope; NTT: New Technology} \\
\multicolumn{5}{l}{$~~~$ Technology Telescope; Palomar: Palomar Hale Telescope; LDT:} \\
\multicolumn{5}{l}{$~~~$ Lowell Discovery Telescope (in 2$\times$2 binning mode)} \\
\multicolumn{5}{l}{$^b$ Field of view} \\
\end{tabular}
\label{table:instrumentation}
\end{table*}

NIRCam observations of 457P were conducted using the F200W and F277W broadband filters, obtained simultaneously with a dichroic and two separate detectors. Each detector has a field of view $2040\times2048$ pixels in size.  The pixel scales are different, however,  $0\farcs0313$~pixel$^{-1}$ and $0\farcs0630$~pixel$^{-1}$ for the F200W and F277W images, respectively, for an angular field of view of $63\farcs9\times64\farcs1$ for F200W images and $128\farcs5\times129\farcs0$ for F277W images.  For the Solar spectrum as reported by \citet{wilmer2018_solarspectrum}, the F200W and F277W filters (covering wavelengths between 1.725~$\mu$m to 2.260~$\mu$m, and between 2.367~$\mu$m to 3.220~$\mu$m, respectively) obtain effective wavelengths of 1.97~$\mu$m and 2.74~$\mu$m, respectively \citep{kelley2023_jwst238p}.  Each of four exposures was obtained using the 4-point INTRAMODULEBOX dither pattern to mitigate detector artifacts and any cosmic ray strikes. The SHALLOW4 readout pattern was used for each exposure, leading to a total exposure time of 1031~s for each filter.

NIRSpec observations of both 457P and an off-source background field were acquired using the NIRSpec integral field unit (IFU) to provide spatially resolved imaging spectroscopy over a $3''\times3''$ field of view. The resulting IFU data cube has spatial elements that are $0\farcs1\times0\farcs1$ in size.  Our observations made use of the IFU's PRISM/CLEAR mode, which provieds a nominal resolving power of $30-330$ between 0.6~$\mu$m -- 5.3~$\mu$m \citep{boker2023_nirspec}. We used the 4-POINT-DITHER dither pattern to obtain four exposures and read out the detecgtor using the NRSIRS2RAPID readout pattern for a total IFU exposure time of 2976~s.

Supporting ground-based observations were obtained with
the Inamori Magellan Areal Camera and Spectrograph \citep[IMACS;][]{dressler2011_imacs} on the 6.5~m Magellan-Baade telescope at Las Campanas in Chile;
the Gemini Multi-Object Spectrograph - North \citep[GMOS-N;][]{hook2004_gmos} in imaging mode on the 8.1~m \rev{Frederick C. Gillett} Gemini North (Gemini-N) telescope (program GN-2024B-Q-114) on Maunakea in Hawaii, USA;
the Gemini Multi-Object Spectrograph - South \citep[GMOS-S;][]{gimeno2016_gmoss} in imaging mode on the 8.1~m Gemini South (Gemini-S) telescope (programs GS-2023A-LP-104, GS-2024A-Q-111, and GS-2024B-Q-113) at Cerro Pach{\'o}n in Chile;
the Large Monolithic Imager \citep[LMI;][]{bida2014_dct} on the 4.3~m Lowell Discovery Telescope (LDT) at Happy Jack, Arizona, USA;
and
the European Southern Observatory (ESO) Faint Object Spectrograph and Camera \citep[EFOSC2;][]{buzzoni1984_efosc} on ESO's
3.58~m New Technology Telescope (NTT; program 113.26J9.002) at La Silla in Chile.
All ground-based observations reported here were obtained using Sloan $r'$-band filters.
Details of all instrumentation are shown in Table~\ref{table:instrumentation}, while observational circumstances of all ground-based observations are listed in Table~\ref{table:ground_observations_457p}.  During several of these attempts to observe 457P, the object was either too faint to detect with the given telescope or could not be clearly identified due to dense background star fields.  For completeness, these observations are still listed in Table~\ref{table:ground_observations_457p} with notes indicating that the object was not detected.

\setlength{\tabcolsep}{6.5pt}
\setlength{\extrarowheight}{0em}
\begin{table*}[htb!]
\caption{Ground-based $r'$-band Observations of 457P$^a$}
\centering
\smallskip
\footnotesize
\begin{tabular}{lcrrcrrrrrrr}
\hline\hline
\multicolumn{1}{c}{UT Date}
 & \multicolumn{1}{c}{Telescope$^b$}
 & \multicolumn{1}{c}{$N$$^c$}
 & \multicolumn{1}{c}{$t$$^d$}
 & \multicolumn{1}{c}{$\theta_s$$^e$}
 & \multicolumn{1}{c}{$\nu$$^f$}
 & \multicolumn{1}{c}{$r_h$$^g$}
 & \multicolumn{1}{c}{$\Delta$$^h$}
 & \multicolumn{1}{c}{$\alpha$$^i$}
 & \multicolumn{1}{c}{PA$_{-\odot}$$^j$}
 & \multicolumn{1}{c}{PA$_{-v}$$^k$}
 & \multicolumn{1}{c}{$\Delta t_q$$^l$}
 \\
\hline
2023 Feb 16 & Gemini-S & 3 & 900 & 0.9 & 223.6 & 2.854 & 1.869 & 2.0 & 63.0 & 284.6 & $-$551 \\ 
2023 Mar 16 & Gemini-S & 3 & 900 & 0.8 & 229.2 & 2.828 & 1.976 & 12.4 & 103.1 & 283.4 & $-$523 \\  
2023 Apr 10 & Gemini-S & 3 & 900 & 0.7 & 234.2 & 2.803 & 2.210 & 18.6 & 105.9 & 282.7 & $-$498 \\  
2023 Apr 22$^*$ & Palomar  & 6 & 1800 & 1.3 & 236.7 & 2.791 & 2.351 & 20.3 & 106.8 & 282.7 & $-$486 \\  
2023 May 08 & Gemini-S & 3 & 900 & 0.9 & 240.1 & 2.773 & 2.556 & 21.3 & 108.0 & 283.0 & $-$469 \\  
2024 Feb 11 & Gemini-S & 6 & 900 & 0.5 & 305.8 & 2.440 & 2.539 & 22.8 & 279.9 & 276.8 & $-$191 \\  
2024 Mar 07$^*$ & Magellan & 4 & 930 & 0.8 & 312.5 & 2.415 & 2.205 & 24.3 & 275.4 & 274.4 & $-$166 \\  
2024 Mar 09 & Gemini-S & 4 & 600 & 0.8 & 313.1 & 2.413 & 2.180 & 24.3 & 275.0 & 274.3 & $-$164 \\  
2024 Apr 04 & Gemini-S & 6 & 900 & 0.9 & 320.2 & 2.391 & 1.844 & 23.0 & 270.3 & 272.8 & $-$138 \\  
2024 May 08$^*$ & LDT      & 16 & 1600 & 1.4 & 329.9 & 2.366 & 1.487 & 15.2 & 261.6 & 272.5 & $-$104 \\  
2024 May 11$^*$ & LDT      & 7 & 700 & 1.3 & 330.7 & 2.364 & 1.464 & 14.2 & 260.2 & 272.6 & $-$101 \\  
2024 Jun 01 & LDT      & 15 & 1500 & 1.3 & 336.7 & 2.353 & 1.355 & 5.9 & 233.7 & 273.2 & $-$80 \\  
2024 Jun 06 & LDT      & 2 & 200 & 1.3 & 338.2 & 2.350 & 1.344 & 4.4 & 211.2 & 273.4 & $-$75 \\  
2024 Jun 07 & Gemini-S & 6 & 900 & 0.6 & 338.4 & 2.350 & 1.343 & 4.2 & 206.2 & 273.4 & $-$74 \\  
2024 Jun 29 & Gemini-S & 18 & 1530 & 0.8 & 344.7 & 2.341 & 1.373 & 9.9 & 117.1 & 274.1 & $-$52 \\  
2024 Aug 02 & Gemini-N & 8 & 720 & 0.7 & 354.7 & 2.333 & 1.617 & 21.3 & 101.5 & 274.6 & $-$18 \\  
2024 Aug 06 & Gemini-N & 8 & 720 & 0.8 & 355.9 & 2.333 & 1.656 & 22.2 & 100.7 & 274.5 & $-$14 \\  
2024 Aug 08 & NTT      & 3 & 900 & 1.2 & 356.4 & 2.333 & 1.673 & 22.5 & 100.3 & 274.5 & $-$12 \\  
2024 Aug 09$^*$ & NTT      & 2 & 600 & 2.2 & 356.7 & 2.333 & 1.684 & 22.7 & 100.1 & 274.5 & $-$11 \\  
2024 Aug 10 & NTT      & 4 & 750 & 1.1 & 357.0 & 2.333 & 1.694 & 22.8 & 99.9 & 274.5 & $-$10 \\  
2024 Aug 11 & NTT      & 14 & 3400 & 0.9 & 357.3 & 2.332 & 1.705 & 23.0 & 99.7 & 274.5 & $-$9 \\  
2024 Sep 23$^*$ & LDT      & 20 & 1200 & 1.1 &  9.9 & 2.336 & 2.204 & 25.3 & 91.7 & 272.1 & $+$34 \\  
2024 Sep 28 & Gemini-S & 3 & 99 & 0.8 & 11.3 & 2.337 & 2.264 & 25.1 & 90.7 & 271.7 & $+$40 \\  
2024 Oct 02-03$^*$ & Magellan & 15 & 675 & 0.7 & 12.8 & 2.338 & 2.324 & 24.8 & 89.7 & 271.1 & $+$44 \\  
2024 Oct 03-04 & Gemini-S & 3 & 99 & 0.6 & 13.0 & 2.339 & 2.335 & 24.7 & 89.5 & 271.0 & $+$44 \\  
2024 Oct 27 & Gemini-S & 8 & 720 & 0.9 & 19.7 & 2.347 & 2.604 & 22.4 & 84.9 & 268.3 & $+$68 \\  
\hline
\hline
\multicolumn{12}{l}{$^a$ Observing geometry parameters from JPL Horizons (Orbit solution \#12) \citep{giorgini1996_horizons}} \\
\multicolumn{12}{l}{$^b$ See Table~\ref{table:instrumentation} for explanations of telescope designations.} \\
\multicolumn{12}{l}{$^c$ Number of usable exposures.} \\
\multicolumn{12}{l}{$^d$ Total exposure time of usable exposures, in s.} \\
\multicolumn{12}{l}{$^e$ FWHM seeing, in arcseconds.} \\
\multicolumn{12}{l}{$^f$ True anomaly, in degrees.} \\
\multicolumn{12}{l}{$^g$ Heliocentric distance, in au.} \\
\multicolumn{12}{l}{$^h$ Geocentric distance, in au.} \\
\multicolumn{12}{l}{$^i$ Solar phase angle (observer-target-Sun), in degrees.} \\
\multicolumn{12}{l}{$^j$ Position angle of the anti-Solar vector as projected on the sky, in degrees East of North.} \\
\multicolumn{12}{l}{$^k$ Position angle of the negative heliocentric velocity vector as projected on the sky, in degrees East of North.} \\
\multicolumn{12}{l}{$^l$ Time prior to (negative values) or after (positive values) perihelion, in days.} \\
\multicolumn{12}{l}{$^*$ Object not clearly identifiable.} \\
\end{tabular}
\label{table:ground_observations_457p}
\end{table*}

\setlength{\tabcolsep}{7.0pt}
\setlength{\extrarowheight}{0em}
\begin{table*}[htb!]
\caption{$r'$-band Photometric Results for 457P (Inactive)}
\centering
\smallskip
\footnotesize
\begin{tabular}{lccc}
\hline\hline
\multicolumn{1}{c}{UT Date}
 & \multicolumn{1}{c}{Telescope$^a$}
 & \multicolumn{1}{c}{$m_{r'}(r_h,\Delta,\alpha)$$^b$}
 & \multicolumn{1}{c}{$m_{r'}(1,1,\alpha)$$^c$}
 \\
\hline
2023 Feb 16 & Gemini-S & 22.27$\pm$0.05 & 18.63$\pm$0.05 \\
2023 Mar 16 & Gemini-S & 23.57$\pm$0.05 & 19.83$\pm$0.05 \\
2023 Apr 10 & Gemini-S & 24.4$\pm$0.1   & 20.4$\pm$0.1 \\
2023 May 08 & Gemini-S & 24.5$\pm$0.2   & 20.2$\pm$0.2 \\
\hline
\hline
\multicolumn{4}{l}{$^a$ See Table~\ref{table:instrumentation} for explanations of telescope designations.} \\
\multicolumn{4}{l}{$^b$ Mean $r'$-band apparent magnitude using photometry apertures } \\
\multicolumn{4}{l}{$~~~~$with radii determined from a curve of growth analysis.} \\
\multicolumn{4}{l}{$^c$ Mean $r'$-band reduced magnitude (normalized to $r_h=\Delta=1$~au).} \\
\end{tabular}
\label{table:ground_photometry_457p_inactive}
\end{table*}

\setlength{\tabcolsep}{6.0pt}
\setlength{\extrarowheight}{0em}
\begin{table*}[htb!]
\caption{$r'$-band Photometric Results for 457P (Potentially Active)}
\centering
\smallskip
\footnotesize
\begin{tabular}{lccccccccc}
\hline\hline
\multicolumn{1}{c}{UT Date}
 & \multicolumn{1}{c}{Telescope$^a$}
 & \multicolumn{1}{c}{$\rho_{0}$$^b$}
 & \multicolumn{1}{c}{$d_{0}$$^c$}
 & \multicolumn{1}{c}{$m_{r'}(r_h,\Delta,\alpha)$$^d$}
 & \multicolumn{1}{c}{$m_{V}(1,1,0)$$^e$}
 & \multicolumn{1}{c}{$A_d/A_n$$^f$}
 & \multicolumn{1}{c}{$Af\rho_{0}$$^g$}
 & \multicolumn{1}{c}{$\rho_{5000{\rm km}}$$^h$}
 & \multicolumn{1}{c}{$Af\rho_{5000{\rm km}}$$^i$}
 \\
\hline
2024 Feb 11$^*$ & Gemini-S & 1.1 & 2000 & 24.2$\pm$0.2 & 18.6$\pm$0.2 & 0.05$\pm$0.13 & 0.1$\pm$0.4 & 2.7 & 0.1$\pm$0.4\\
2024 Mar 09$^*$ & Gemini-S & 1.1 & 1700 & 23.9$\pm$0.3 & 18.6$\pm$0.2 & 0.08$\pm$0.17 & 0.3$\pm$0.6 & 3.2 & 0.2$\pm$0.6 \\
2024 Apr 04 & Gemini-S & 1.0 & 1300 & 23.6$\pm$0.3 & 18.7$\pm$0.4 & 0.0 & 0.0 & 3.7 & 0.0 \\
2024 Jun 01 & LDT      & 1.6 & 1600 & 21.8$\pm$0.3 & 18.7$\pm$0.3 & 0.0 & 0.0 & 5.1 & 0.0 \\
2024 Jun 06 & LDT      & 1.4 & 1400 & 21.66$\pm$0.01 & 18.73$\pm$0.06 & 0.0 & 0.0 & 5.1 & 0.0 \\
2024 Jun 07 & Gemini-S & 1.1 & 1100 & 21.78$\pm$0.01 & 18.87$\pm$0.06 & 0.0 & 0.0 & 5.1 & 0.0 \\
2024 Jun 29 & Gemini-S & 1.6 & 1600 & 22.00$\pm$0.09 & 18.6$\pm$0.1 & 0.05$\pm$0.09 & 0.2$\pm$0.3 & 5.0 & 0.1$\pm$0.3 \\
2024 Aug 02 & Gemini-N & 3.2 & 3800 & 22.1$\pm$0.1 & 18.1$\pm$0.1 & 0.7$\pm$0.2 & 1.1$\pm$0.2 & 4.3 & 0.9$\pm$0.4 \\
2024 Aug 06 & Gemini-N & 3.2 & 3800 & 22.14$\pm$0.08 & 18.1$\pm$0.1 & 0.7$\pm$0.2 & 1.1$\pm$0.2 & 4.2 & 1.0$\pm$0.4 \\
2024 Aug 08 & NTT      & 1.7 & 2100 & 22.2$\pm$0.3 & 18.1$\pm$0.2 & 0.7$\pm$0.4 & 2.1$\pm$1.1 & 4.1 & 1.2$\pm$1.1 \\
2024 Aug 10 & NTT      & 1.9 & 2300 & 22.5$\pm$0.1 & 18.3$\pm$0.1 & 0.4$\pm$0.1 & 1.2$\pm$0.3 & 4.1 & 0.7$\pm$0.3 \\
2024 Aug 11 & NTT      & 2.2 & 2700 & 22.1$\pm$0.2 & 18.0$\pm$0.2 & 0.9$\pm$0.3 & 2.0$\pm$0.6 & 4.0 & 1.4$\pm$0.8 \\
2024 Sep 28 & Gemini-S & 0.6 & 1100 & 23.2$\pm$0.1 & 18.3$\pm$0.1 & 0.4$\pm$0.1 & 2.3$\pm$0.7 & 3.1 & 0.9$\pm$0.7 \\
2024 Oct 03-04 & Gemini-S & 0.8 & 1400 & 23.6$\pm$0.1 & 18.5$\pm$0.2 & 0.3$\pm$0.1 & 1.1$\pm$0.5 & 3.0 & 0.5$\pm$0.5 \\
2024 Oct 27 & Gemini-S & 1.0 & 1900 & 23.5$\pm$0.1 & 18.3$\pm$0.1 & 0.4$\pm$0.1 & 1.2$\pm$0.3 & 2.7 & 0.7$\pm$0.4 \\
\hline
\hline
\multicolumn{10}{l}{$^a$ See Table~\ref{table:instrumentation} for explanations of telescope designations.} \\
\multicolumn{10}{l}{$^b$ Aperture radius, in arcseconds, determined to be optimal for photometry measurements from a curve of growth analysis.} \\
\multicolumn{10}{l}{$^c$ Distance, in km, equivalent to $\rho_0$ at the geocentric distance of the comet.} \\
\multicolumn{10}{l}{$^d$ Mean $r'$-band apparent magnitude using photometry apertures with $\rho_0$.} \\
\multicolumn{10}{l}{$^e$ Mean $V$-band absolute magnitude (normalized to $r_h=\Delta=1$~au and $\alpha=0^{\circ}$) using photometry apertures with $\rho_0$.} \\
\multicolumn{10}{l}{$^f$ Inferred dust-to-nucleus scattering surface area ratio} \\
\multicolumn{10}{l}{$^g$ $Af\rho$, in cm, computed from photometry measured using apertures with $\rho_0$.} \\
\multicolumn{10}{l}{$^h$ Angular equivalent, in arcseconds, to a $\rho=5000$~km photometry aperture at the geocentric distance of the comet.} \\
\multicolumn{10}{l}{$^i$ $Af\rho$, in cm, for $\rho=5000$~km computed from a best-fit power law to measured $Af\rho$ vs.\ aperture data, with uncertainties } \\
\multicolumn{10}{l}{$~~~~$incorporating the range of $Af\rho$ values for $\rho=(5000\pm2000)$~km.} \\
\multicolumn{10}{l}{$^*$ Not clearly visibly active; nominal photometric excess with respect to computed absolute nucleus magnitude } \\
\multicolumn{10}{l}{$~~~~$considered potentially due to rotational variation.} \\
\end{tabular}
\label{table:ground_photometry_457p_active}
\end{table*}

All ground-based observations were conducted using non-sidereal tracking and at airmasses of $\lesssim2.5$, with typical seeing conditions of $\sim1''-2''$.  A minimum of three exposures was obtained during each visit in order to ensure that our target and any associated activity could be unambiguously identified from their non-sidereal motion.  In some cases, however, certain detections in a sequence were discarded due to being too close to background sources for photometry to be reliable, leading to fewer than 3 exposures per night being reported here.

\section{Data Reduction\label{section:data_reduction}}

\subsection{\jwst{} NIRSpec Data Reduction\label{section:nirspec_data_reduction}}

To properly account for any background emissions in our NIRSpec observations of 457P, background observations were acquired from a position 1\arcmin{} away from the target. These separate background points are necessary as the extended nature of comets does not necessarily mean that a gas- and/or dust-free background spectrum can be obtained within the $3\arcsec\times3\arcsec$ FOV of the on-source data. To correct for any contamination from streaked background stars in several groups the {\tt jump} step was used in the {\tt detector1} pipeline, as was a custom step to extend the jump mask, in \jwst{} {\tt python} development version 1.21.0. The background was removed using the \jwst{} {\tt python} package and the {\tt spec2} data reduction pipeline\footnote{\url{https://jwst-pipeline.readthedocs.io/}} version 1.16 \citep{bushouse2023_jwstcalibrationpipeline_1_12_5}.  In particular, the NSClean step was used to reduce detector $1/f$ noise \citep{rauscher2024_nsclean}.
At this point in the pipeline, the data were photometrically calibrated to radiance units of MJy~sr$^{-1}$. 

\begin{figure*}
    \centering
    \includegraphics[width=0.49\linewidth]{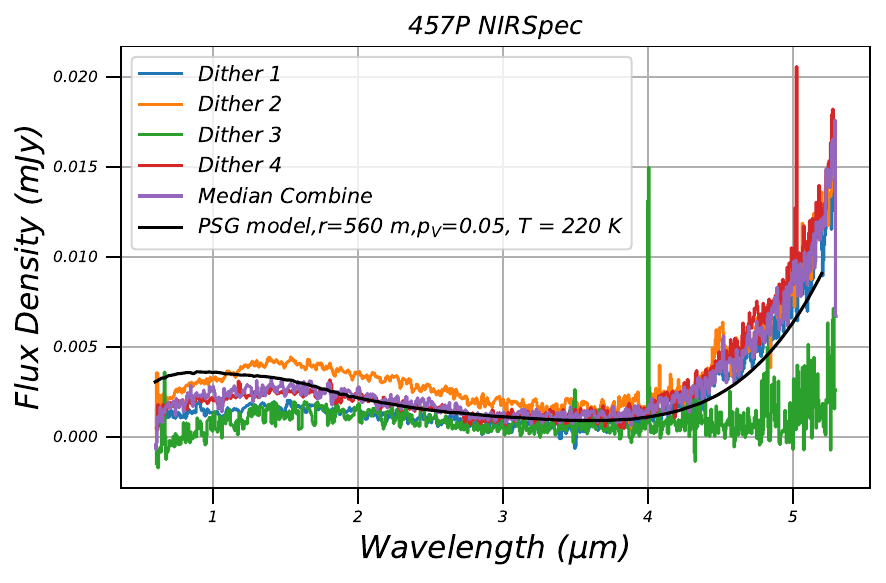}
    \includegraphics[width=0.49\linewidth]{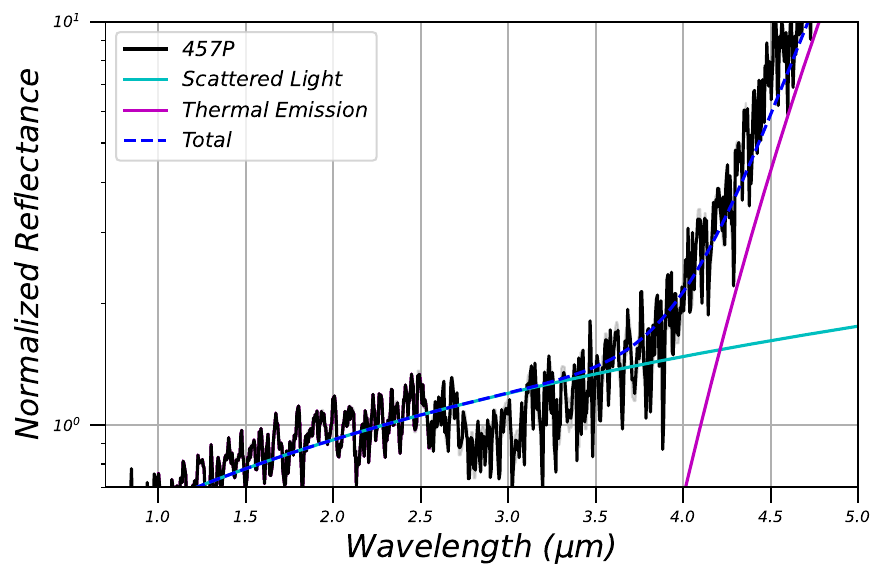}
    \caption{Left: NIRSpec Extracted Spectrum from 1.8 to 5.2 microns for each of our four dithers, compared to our initial Planetary Spectrum Generator model estimate for a bare $p_V$=0.05, T = 220 K, $r=560 m$, nucleus at our NIRSpec observing geometry to illustrate the weak dust contribution. Due to the poor signal in Dither 3 we exclude it from our median combined spectrum that is used for analysis. Note that the variation between the dithers is largely than the $\sim$25-30\% variation expected based on HST lightcurve measurements \citep{kim2022_p2020o1}. Right: Results from our normalized reflectance modeling of H$_2$O, with our normalized  median combined spectrum divided by the solar spectrum in black, our linear reflectance in cyan, a simple thermal model for a 180 K black body normalized at 4.1 $\mu$m in magenta, and the total continuum model shown with a dotted blue line.}
    \label{fig:457P_NIRSpec}
\end{figure*}

Following the pipeline steps we extract a one-dimensional spectrum from each dither exposure with a $0\farcs4$ radius aperture (Figure~\ref{fig:457P_NIRSpec}). By implementing a slice-by-slice method in {\tt python} centered on the brightest pixel near the comet's predicted location in the frame we are able to mitigate the effect of small pixel-scale deviations that may occurred for the optocenter  as a function of wavelength, though these changes are expected to be in the tenth of a pixel range. To prevent any confusion of the optocenter calculation by the ``snowball'' cosmic ray events \citep{regan2024_snowballs}, which are infrequently missed during the prior cosmic ray cleaning algorithms in the \jwst{} {\tt python} steps, an additional algorithm locates all pixels $50\sigma$ above the median signal of their nearest neighbors. These flagged pixels then had their values replaced with the median of their nearest neighbors which allows us to find the extended cosmic ray events without removing excess dust coma signal. We note that the third dither does not show the signal-to-noise of the other three (see Fig.~\ref{fig:457P_NIRSpec}, left panel), and therefore we have omitted it from our median combined spectrum. Following these two custom steps, we found that the one-dimensional spectra extracted from the dataset vary in near-infrared (NIR) flux by a factor of $\sim$3, while the thermal flux appears constant. This may suggest variations in NIR albedo across the surface of 457P, but individual dithers do not have sufficient signal to constrain this further. This is notable given that \citet{kim2022_p2020o1} found that 457P's lightcurve only varied by 0.3 mag, or $\sim24\%$. 

\begin{figure*}[ht]
    \centering
    \includegraphics[width=0.8\linewidth]{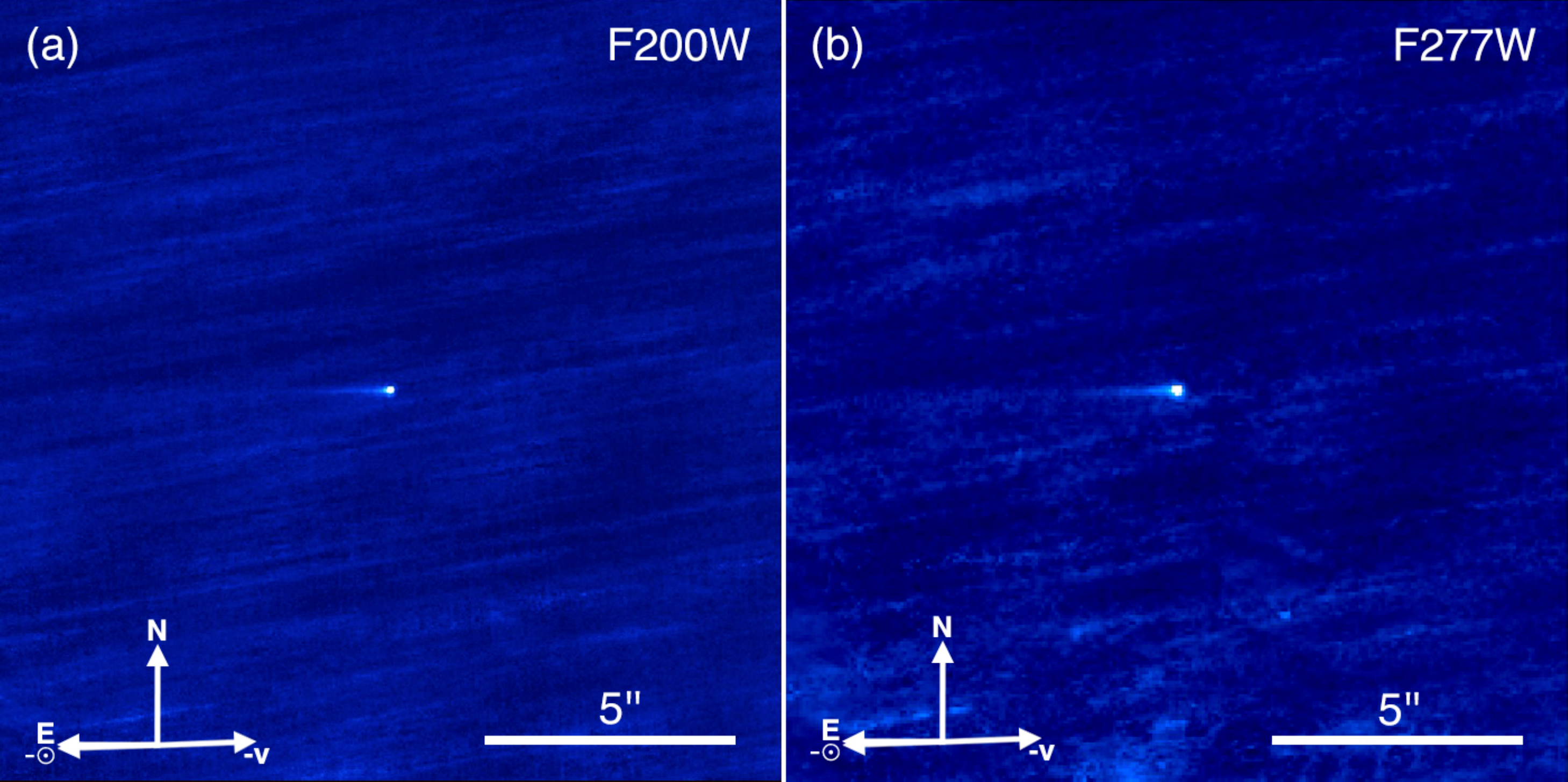}
    \caption{Median composite images of 457P/Lemmon-PANSTARRS, aligned on the photocenter of the comet in each individual image, constructed from NIRCam data obtained using the (a) F200W and (b) F277W broadband filters, comprising 1031~s of total exposure time each.  Labeled arrows indicate the directions of celestial north (N) and east (E), and the projected anti-Sun ($-\odot$) and negative heliocentric velocity ($-v$) vectors as seen from \jwst{}. A $5''$ angular scale bar (7900 km at the distance of the comet) is also shown in each panel. Color scaling in both panels is linear, where regions in the inner coma (in the center of each image) that are shown as solid white consist of pixels with fluxes that are $\sim25$\% of the peak central flux or larger.}
    \label{fig:nircam_images}
\end{figure*}

\subsection{\jwst{} NIRCam Data Reduction\label{section:nircam_data_reduction}}

NIRCam images were processed with pipeline version 12.0.5 and Calibration Reference Data System context file number 1303.  
Following \citet{hsieh2025_358p},
we performed additional processing to mitigate the effects on subsequent processing of visible horizontal ``striped'' background structure in individual images attributed to $1/f$ noise from \jwst{}'s SIDECAR ASICs detector readout electronics\footnote{\url{https://jwst-docs.stsci.edu/known-issues-with-jwst-data/nircam-known-issues/nircam-1-f-noise-removal-methods}}, as well as relatively large numbers of randomly distributed pixels with {\tt NaN} values.

For this additional processing, we first used the {\tt image1overf} package developed by C.\ Willott\footnote{\url{https://github.com/chriswillott/jwst}} to remove row median levels while preserving overall flux levels in order to reduce the $1/f$ background structure.
We then replaced {\tt NaN} pixel values (which comprised $\sim1$\% of all pixels in the uncorrected images) and uncertainties with the median value and uncertainty of all adjacent non-{\tt NaN} pixels using our own {\tt python} code.

To better study the detailed morphology of the comet and mitigate the effect of cosmic rays and detector artifacts, we then used {\tt pyraf}\footnote{\url{https://pypi.org/project/pyraf/}} \citep{stsci2012_pyraf} to construct median composite images of the object in each filter by shifting and aligning individual images on the object's photocenter using linear interpolation and performing a median combination of the resulting images for each filter.  Final median composite images are shown in Figure~\ref{fig:nircam_images}.

\subsection{Optical Data Reduction\label{section:optical_data_reduction}}

Standard bias subtraction, flatfield correction, and cosmic ray removal were performed for all optical images obtained from ground-based facilities using {\tt python} code utilizing the {\tt ccdproc} package\footnote{\url{https://ccdproc.readthedocs.io/}} \citep{craig2023_ccdproc} in {\tt astropy}\footnote{\url{http://www.astropy.org}} \citep{astropy2018_astropy} and the {\tt L.A.Cosmic} {\tt python} module\footnote{\url{https://pypi.org/project/lacosmic/}} \citep{vandokkum2001_lacosmic,vandokkum2012_lacosmic}.
Photometry measurements of the target object and at least one background reference star were performed using {\tt IRAF} \citep{tody1986_iraf,tody1993_iraf,fitzpatrick2024_iraf} and {\tt pyraf} software,
where photometry of reference stars was obtained by measuring net fluxes within circular apertures with sizes chosen using curve-of-growth analyses of representative stars, with background sampled from surrounding circular annuli.
Photometry of target objects was performed using circular apertures, where background statistics were measured in nearby but non-adjacent regions of blank sky to avoid potential flux contamination from any coma or nearby field stars.

Absolute photometric calibration was performed using field star magnitudes from the Asteroid Terrestrial-impact Last Alert System \citep[ATLAS;][]{tonry2011_atlas,tonry2018_atlas} Refcat2 all-sky stellar reference catalog \citep{tonry2018_refcat}, where conversion of non-SDSS photometry to magnitudes in the SDSS system was accomplished as needed using transformations derived by \citet{tonry2012_ps1} and by R.~Lupton\footnote{{\url{http://www.sdss.org/}}}. 

We aimed to use up to $10-20$ well-isolated reference stars (i.e., field stars with no other neighboring sources within the photometry aperture used for those data, and ideally, within the annuli used to measure sky background as well) for photometric calibration where possible.
In some cases, however, only a few suitable reference stars, or even just one, were available due to the small margin between the limiting magnitude of the Refcat2 catalog and the saturation limit of many of our observations.
In the cases of Gemini-S observations on UT 2023 February 16 and UT 2023 March 16, no available Refcat2 sources in the field were unsaturated, and so we added an intermediate step in our photometric calibration process where we determined the magnitudes of fainter stars in the fields from Dark Energy Camera (DECam) Legacy Survey observations of the same fields\footnote{Obtained on UT 2015 Apr 11 (PI: D.\ Schlegel; Observers: A.\ Dey, B.\ Nord, B.\ Blum; Proposal ID 2014B-0404) and UT 2016 Jan 11 (PI: D. Schlegel; Observers: Beaudin, Moustakas, Blum; Proposal ID 2014B-0404) for the UT 2023 Feb 16 and UT 2023 Mar 16 Gemini-S fields, respectively}, and then used those to calibrate the Gemini-S data.  This process took advantage of the larger fields of view and shallower image depths of the DECam Legacy Survey data relative to the Gemini-S data, making unsaturated Refcat2 sources available, which could then be used to determine calibrated magnitudes of fainter stars in the field which were below the saturation limit in the Gemini-S data.

To characterize the comet's intrinsic brightness evolution, calibrated photometry were normalized to $r_h=\Delta=1$~au (producing reduced magnitudes) and then to a Solar phase angle of $\alpha=0^{\circ}$ to produce equivalent absolute magnitudes. Reduced magnitudes were computed by applying a correction of $-5\log (r_h\Delta)$ to the measured apparent magnitude, while absolute magnitudes were computed by determining the nucleus's reduced magnitude at the value of $\alpha$ at the time of observations using the IAU phase function parameters computed in Section~\ref{section:nucleus_phase_function}, subtracting the nucleus's reduced magnitude from the total measured reduced magnitude of the comet (i.e., leaving the reduced magnitude of the dust), correcting the reduced magnitude of the dust to $\alpha=0^{\circ}$ using the Schleicher-Marcus phase function\footnote{\url{https://asteroid.lowell.edu/comet/dustphase.html}} \citep[sometimes also referred to as the Halley-Marcus phase function;][]{schleicher2011_sw3,schleicher1998_halley,marcus2007_cometphasefunction} to obtain the effective absolute magnitude of the dust, $m_d(1,1,0)$, and finally adding back the absolute magnitude of the nucleus ($H_V$; see Section~\ref{section:nucleus_phase_function}).  These absolute magnitude results (also converted to $V$-band, assuming solar colors) are shown in Table~\ref{table:ground_photometry_457p_active}.
As part of these calculations, we are also able to compute the ratio of scattering cross-sections of the dust and the nucleus, $A_d/A_n$, where
\begin{equation}
    {A_d\over A_n} = 10^{0.4\left[H_V-m_d(1,1,0)\right]}
\end{equation}
where these values are also shown in Table~\ref{table:ground_photometry_457p_active}. We note that these calculations assumed that the dust coma is optically thin. 
These calculations and others described in subsequent sections use the {\tt uncertainties} {\tt python} package for the calculation and propagation of uncertainties\footnote{\url{http://pythonhosted.org/uncertainties/}}. 

\section{Results and Analysis}

\subsection{Spectroscopy}\label{section:spectroscopy}

\subsubsection{Reflectance Spectroscopy}\label{section:refl_spectroscopy}

\begin{figure}
    \centering
    \includegraphics[width=0.95\linewidth]{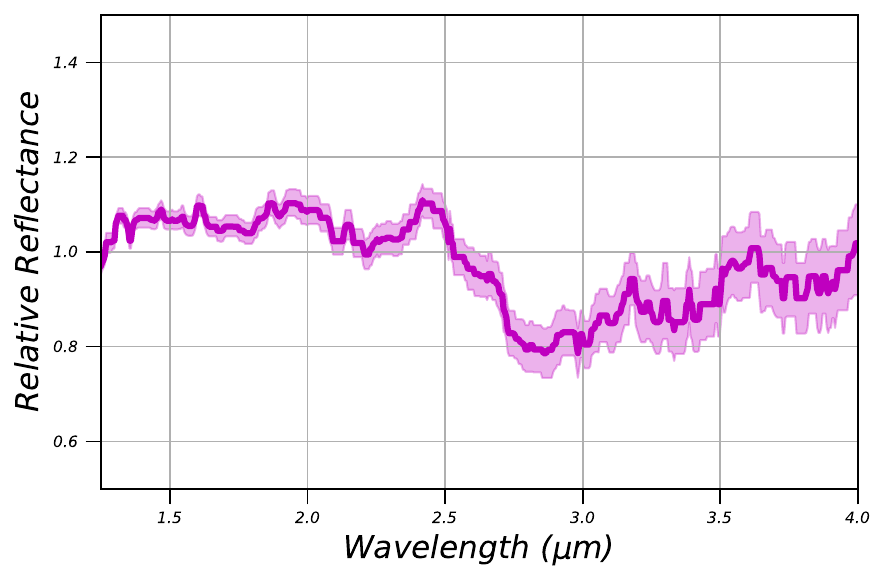}
    \caption{Relative reflectance spectrum for 457P/Lemmon-PANSTARRS after dividing by a solar reference spectrum and a linear continuum to highlight broad absorption bands. The spectrum has been binned by a factor of 35 to highlight the 3 $\mu$m absorption band. Note that the slight upwards trend near 4$\mu$m is due to increased thermal contribution (see Fig. \ref{fig:457P_NIRSpec}, right panel). The broad band width of 457P's 3 $\mu$m absorption is qualitatively similar to 67P/Churyumov-Gerasimenko from VIRTIS \citep{raponi2016_67p,raponi2020_cometorganics}, but lacks the sharp absorption at 2.7 $\mu$m seen on Ceres, Pallas, or Hygeia with JWST \citep{rivkin2025_jwst_mbas}.}
    \label{fig:457P_relative_ref}
\end{figure}

By dividing our extracted one-dimensional spectrum by the STScI CALSPEC Solar reference spectrum \citep{bohlin2022calspec}, we can produce a reflectance spectrum of 457P to search for signs of absorption features due to H$_2$O ice or ammoniated phyllosilicates, or any other diagnostic features in the 3-$\mu$m region that could be useful for comparison to the broader asteroid and cometary populations \citep{rivkin2022_3micron}. We present our Solar-divided reflectance spectrum, normalized at 2~$\mu$m, with a reflectance and 220~K blackbody thermal model fit, in the right panel of Figure~\ref{fig:457P_NIRSpec}. 

Our modeled spectral slope is $S'=(3.79\pm0.01)$\%/100~nm, slightly redder than the slopes reported for 238P/Read and 358P/PANSTARRS $(2.18\pm0.02)$\%/100~nm and $(2.25\pm0.03)$\%/100~nm, respectively \citep{kelley2023_jwst238p,hsieh2025_358p}. All spectra were normalized at 2 $\mu$m.  This is likely due to a greater contribution to the overall spectrum from nucleus reflectance, considering that the phase angles between the targets are all similar, eliminating phase reddening as a possible explanation. \rev{In the absence of any optical, NIR, or IR spectra obtained for these three MBCs when they are inactive near aphelion (and thus also at their faintest when observed from Earth), we must proceed with the caveat that dust reflectance likely contaminates our observed nuclear spectrum to some degree. } For 457P, the contribution of the nucleus' reflectance spectrum to the total observed spectrum \rev{ was estimated with a reflectance/thermal model for a cometary albedo of 0.05, our derived radius of $r_n=0.56$~km for 457P, and a temperature of 220 K (see NASA's Planetary Spectrum Generator model in left panel of Fig. \ref{fig:457P_NIRSpec}, also Section~\ref{section:nucleus_phase_function}). The dust contribution was estimated based on scaling of the 0.6\farcs aperture Gemini-S data obtained on 2024 Sep 28 (Table \ref{table:ground_photometry_457p_active}) to our JWST 0.4\farcs aperture (assuming a 1/$r$ distribution), and modeling the expected reflectance for an Af$\rho$ of 0.92 cm using the PSG. This estimation shows that the nucleus dominates our spectrum by factors of 36-43 }compared to the dust \rev{contribution} for the $Af\rho$ values measured \rev{optically one week after} the JWST observations. 

To match the observed intensity we find that the albedo of 457P is likely slightly higher than the typically assumed $p_V=0.04$ for comets; using the $r_n=0.42$~km for 457P from \citet{kim2022_p2020o1} would require an albedo of $p_V = \sim0.1$. If we use our photometry-derived $r_n=0.56$~km for 457P (see Section~\ref{section:nucleus_phase_function}), and assuming that our spectral slope applies from $V$ to our bandpass, we find that $p_V = 0.05$ produces a better match to our peak observed reflectance flux of 0.003 mJy at $\sim1.5$~$\mu$m, where $\gg99$\% of the observed flux is from scattered sunlight.  

After dividing out the linear continuum, we are then able to produce a relative reflectance spectrum to assess broad absorption features, shown in Figure~\ref{fig:457P_relative_ref}. We find that at the extreme red edge of the absorption feature, 3.3 $\mu$m, there is only about a 2$\%$ contribution from the thermal component of the nucleus and dust, and that our detection of the feature is not influenced within uncertainties by the thermal component.  

\begin{figure}
    \centering
    \includegraphics[width=0.95\linewidth]{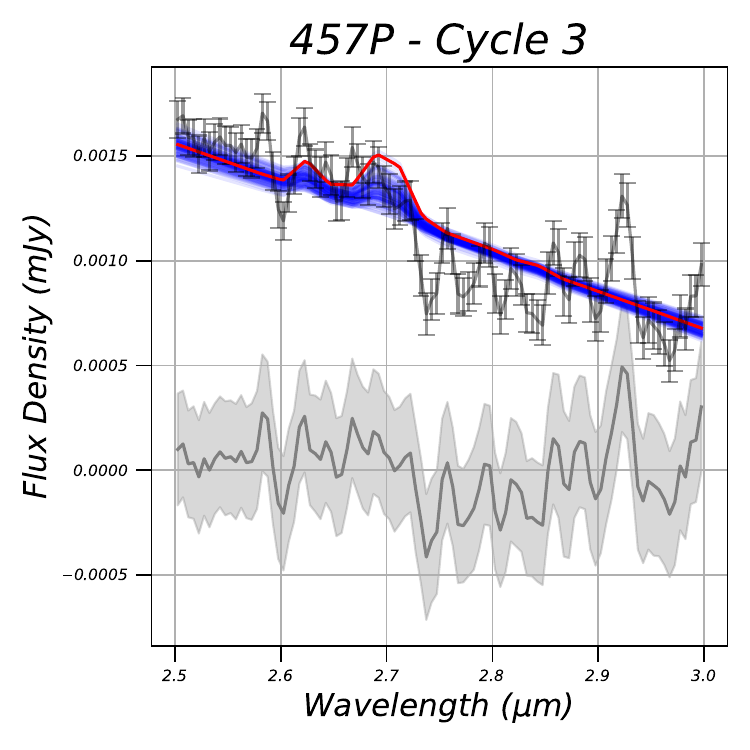}
    \caption{Results from MCMC modeling of H$_2$O, with black spectra showing a subset of the MCMC models, and the red line highlighting our 3$\sigma$ upper limit model for 2.0$\times$10$^{24}$ molecules~s$^{-1}$.}
    \label{fig:457P_h2o_upper}
\end{figure}

Examining our final relative reflectance spectrum, a significant 3 $\mu$m absorption feature is immediately apparent. The width of the feature is at least 1 $\mu$m, with a band center near 3 $\mu$m, and a depth of 20\% relative to the continuum, consistent with the width and band center for the average nonsharp-type spectrum shown in Figure 9 of \citet{rivkin2022_3micron}, a feature that is more prevalent in the outer portions of the main asteroid belt. There also appear to be multiple unique absorption bands at 2.7~$\mu$m, 3.0~$\mu$m, and 3.3~$\mu$m, consistent with  spectra of outer main-belt and cometary objects. We note that the asymmetric shape of the absorption band is substantially different than the sharp-type asymmetric absorption band seen with JWST on similar asteroids \citep{arredondo2024_polana_henrietta,arredondo2025_jwst_polana}, but qualitatively similar to objects like Cybele and 67P/Churyumov-Gerasimenko \citep{rivkin2022_3micron}, \rev{with the exception of the 2.7 $\mu$m feature, which is substantially deeper on 457P}. This comparison to the \rev{sharp-type (ST) and} nonsharp-type (NST) low-albedo asteroids \rev{as well as} the implications are discussed further in Section \ref{sec:457P_in_context}. 

\begin{figure*}
    \centering
    \includegraphics[width=0.32\linewidth]{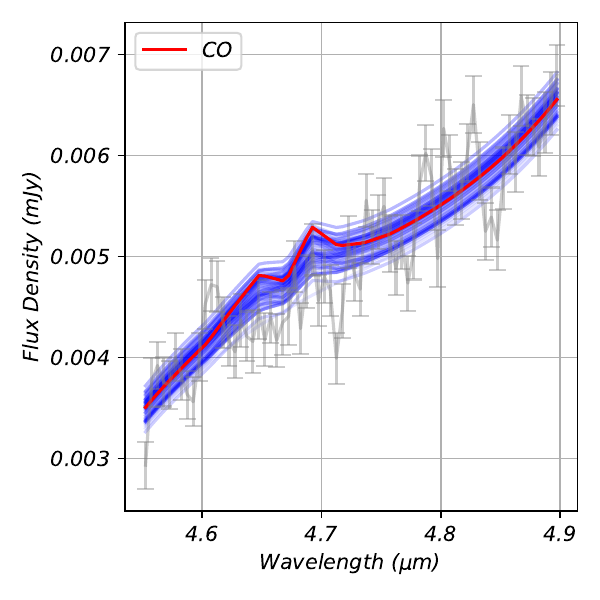}
    \includegraphics[width=0.32\linewidth]{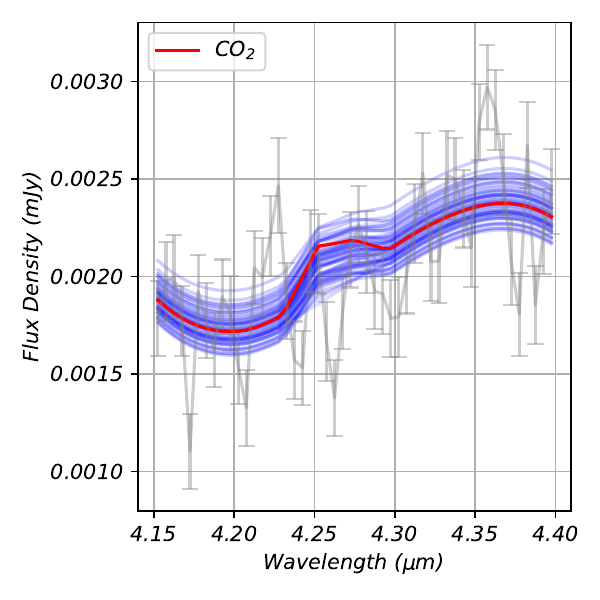}
    \includegraphics[width=0.32\linewidth]{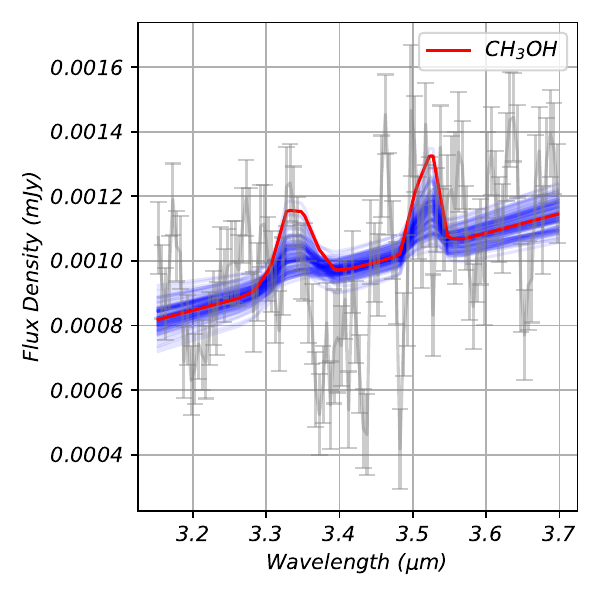}
    \caption{Upper limit fits to JWST NIRSpec data for 457P for CO, CO$_{2}$, and CH$_{3}$OH, with production rates reported in Table \ref{table:jwst_upper_limits}. Data are shown in gray, a subsample of MCMC fits is shown in blue, and the derived 3$\sigma$ upper limit is shown in red. Using polynomial background contributions did not meaningfully change the upper limit determinations.}
    \label{fig:457P_upperlims}
\end{figure*}

\begin{table}[]
    \centering
    \begin{tabular}{c|c}
         Volatile & 3$\sigma$ Upper limit (molec. s$^{-1}$) \\
         \hline
         \hline
         H$_{2}$O & 2.0$\times$10$^{24}$ \\
         CO & 4.8$\times$10$^{24}$\\
         CO$_{2}$ & 1.4$\times$10$^{23}$\\
         CH$_3$OH & 2.2$\times$10$^{24}$\\       
    \end{tabular}
    \caption{3$\sigma$ upper limit determinations from JWST NIRSpec observations of 457P/Lemmon-PANSTARRS on 2024 September 20.}
    \label{table:jwst_upper_limits}
\end{table}

\subsubsection{Volatile Species Upper Limits}\label{section:volatile_species}

We find no evidence for molecular emissions from H$_{2}$O, CO, CO$_{2}$, or CH$_3$OH in our median combined spectrum, with meaningful upper limits that suggest no current detectable sublimation of volatile reservoirs when using a rotational temperature of 35 K, intermediate between the two temperatures of 15 K and 45K implemented in \citet{hsieh2025_358p}. We note that using the lower of the two values would decrease our upper limits by approximately 30\%, while using the higher temperature would increase it by about 10\%. To find rigorous 3$\sigma$ upper limits for production rates, we employed the {\tt emcee} Python package \citep{foremanmackey2013_emcee} for Markov chain Monte Carlo (MCMC) sampling, simultaneously evaluating a fit of emissions and a background continuum to comprehensively determine the maximum volatile production rates that could be masked within the uncertainties. The emission spectrum was retrieved from the NASA Planetary Spectrum Generator \citep{villanueva2018_psg} for the specific observing geometry on UT 2024 September 20 from JWST, and our MCMC model linearly scaled the estimated production rates  combined with slight changes to the continuum fit to empirically characterize our upper limits on production rate. Our 3$\sigma$ upper limits for all four volatiles are reported in Table \ref{table:jwst_upper_limits} and shown in Figures \ref{fig:457P_h2o_upper} and \ref{fig:457P_upperlims}. 

Using the ice sublimation model developed by \citet{cowan1979_cometsublimation} and available at the NASA Planetary Data System's Small Bodies Node\footnote{\url{https://pds-smallbodies.astro.umd.edu/tools/ma-evap/index.shtml}}, we can compute the expected water ice sublimation rate, ${\dot m_w}$, from the surface of of a comet at a given heliocentric distance. 
At $r_h=2.335$~au where 457P was observed by NIRSpec, we find ${\dot m_{w}}\sim 3.9\times10^{20}$~molecules~s$^{-1}$~m$^{-2}$ for a non-rotating or pole-on sphere (which produces the maximum attainable temperature for an object), and ${\dot m_{w}}\sim 1.2\times10^{20}$~molecules~s$^{-1}$~m$^{-2}$ in the zero-obliquity, fast-rotating case.  Given our measured upper limit water production rate of $Q_{\rm H_2O}<2.0\times10^{24}$~molecules~s$^{-1}$, we compute upper limit effective active areas of 
$A_{\rm act}=5130$~m$^2$ in the non-rotating or pole-on case and 
$A_{\rm act}=16670$~m$^2$ in the zero-obliquity, fast-rotating case.  These effective active areas correspond to circular areas $<40$~m and 
$<73$~m in radius, and effective active fractions of
$f_{\rm act}<1.3\times10^{-3}$ and 
$f_{\rm act}<4\times10^{-3}$, assuming a spherical nucleus with $r_n=(0.56\pm0.06)$~km (Section~\ref{section:nucleus_phase_function}).

\subsection{Photometry}\label{section:photometry}

\subsubsection{Nucleus Phase Function}\label{section:nucleus_phase_function}

\begin{figure}
    \centering
    \includegraphics[width=1.00\linewidth]{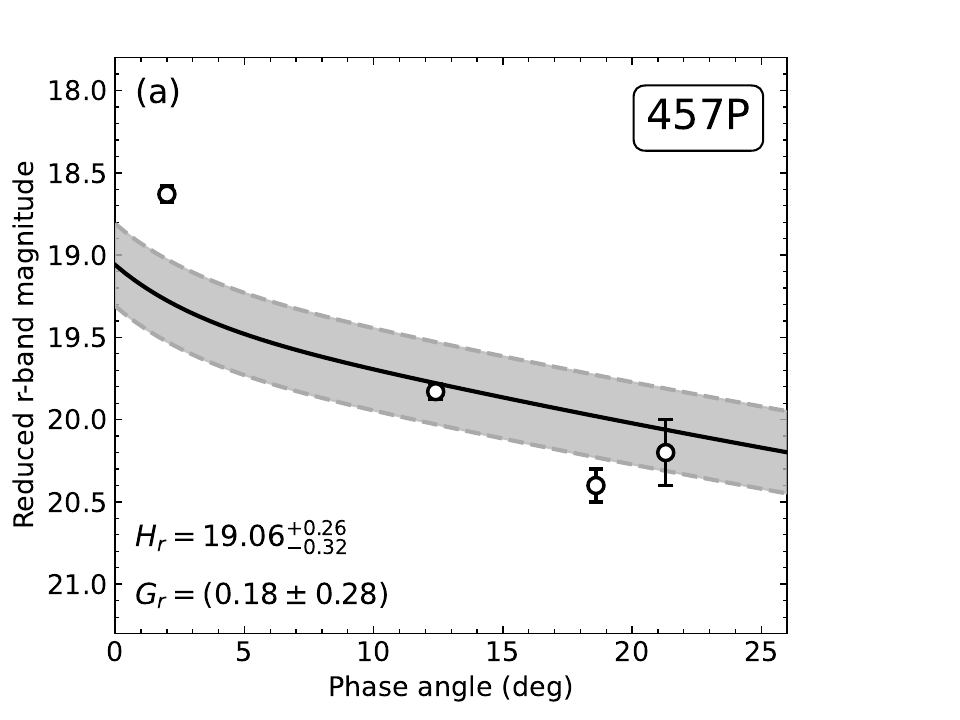}
    \includegraphics[width=1.00\linewidth]{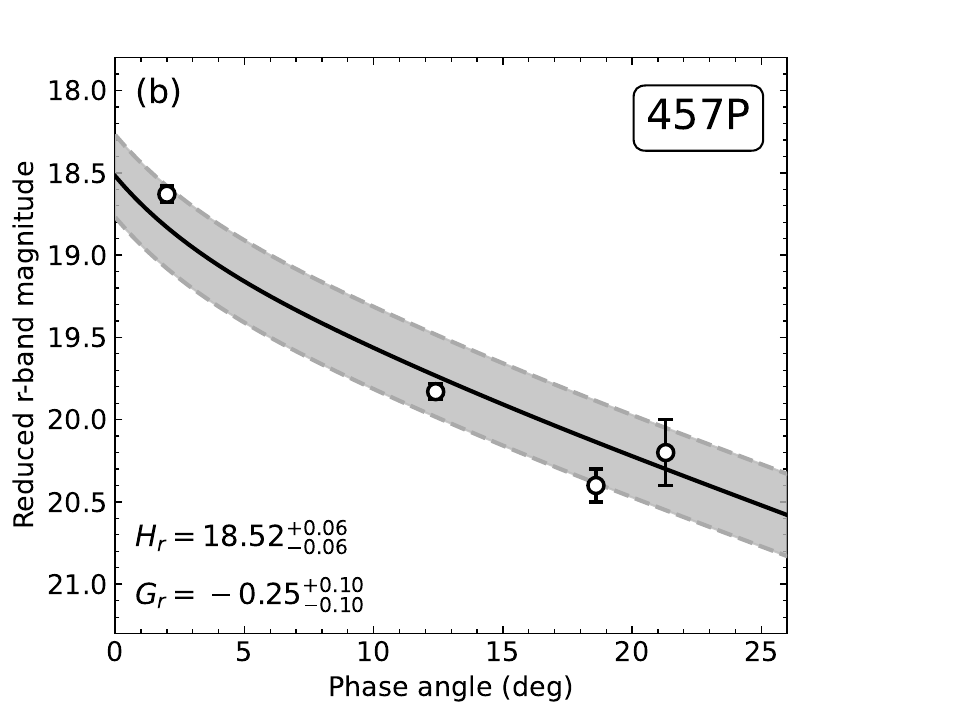}
    \caption{(a) Phase function fit for 457P's inactive nucleus using an assumed value of $G=0.18\pm0.28$. (b) Phase function fit for 457P's inactive nucleus allowing both $H$ and $G$ to vary freely.  In both panels, the gray shaded region shows the range of potential photometric variation for a nucleus with a rotational lightcurve amplitude of $A=0.25$~mag.}
    \label{fig:phase_function}
\end{figure}

In advance of our 2024 \jwst{} observations of 457P and accompanying ground-based observations, we also obtained ground-based optical observations of the target on five occasions in 2023, four of which resulted in successful detections (see Table~\ref{table:ground_observations_457p}).  Since the object was not expected to be active during this time (and no visible activity was seen in the data), these detections provided an opportunity to independently verify the absolute magnitude estimate for the inactive nucleus of $H_V=19.25\pm0.13$ reported by \citet{kim2022_p2020o1}, based on an IAU assumed phase function slope parameter of $G_V=0.15$.

For this analysis, we use the same methods employed by \citet{hsieh2023_mbcnuclei}, where a Monte Carlo-style approach was used to obtain phase function parameter fits with realistic uncertainties for other MBC nuclei.  Using this approach, we generated a large number of individual test data sets based on the photometric data shown in Table~\ref{table:ground_photometry_457p_inactive}, derived the best-fit phase function parameters for each test data set, and then identified the median best-fit values and 1$\sigma$ intervals for those parameter values from the distribution of results from the set of fitting runs.  Test data sets were generated by starting with the original mean magnitudes measured for the object on each night, and applying Gaussian-distributed photometric offsets using 1$\sigma$ values equal to the measured uncertainties of each photometric point.  Least-squares fitting was carried out using the Levenberg-Marquardt algorithm as implemented by the {\tt LevMarLSQFitter} function in {\tt astropy}.  Median values of phase function parameters from the aggregate results of all testing runs were then adopted as nominal solutions, with upper and lower 1$\sigma$ uncertainty intervals determined by identifying the intervals enclosing 34\% of the total results set above and below the computed median value.

Due to the small number of available photometric points, we first carried out our fitting analysis by adopting a fixed $G$ parameter value \citep[similar to the approach used by][]{kim2022_p2020o1}, where we adopted the value of $G_r=0.18\pm0.28$ used by \citet{hsieh2023_mbcnuclei} as an empirically determined typical value for C-type asteroids, finding a best-fit absolute magnitude of $H_r=\left(19.06^{+0.26}_{-0.32}\right)$~mag (Figure~\ref{fig:phase_function}a).  Assuming solar colors \citep{jordi2006_filtertransformations,holmberg2006_solarcolors}, we find this to be equivalent to $H_V=(19.27\pm0.3)$~mag, which is remarkably consistent with the value of $H_V=(19.25\pm0.13$)~mag found by \citet{kim2022_p2020o1}.  Noting the poor fit of this phase function to our relatively high-confidence data point at $\alpha=2.0^\circ$ \citep[which is almost 0.7~mag brighter than expected from the best-fit phase function, well in excess of the 0.3~mag peak-to-peak lightcurve variation of the object reported by][; see Figure~\ref{fig:phase_function}a]{kim2022_p2020o1}, however, we conducted another analysis solving for both $H$ and $G$ (instead of solving only for $H$ and using a fixed assumed $G$ value), finding a much better best-fit solution compatible with all available photometry where $H_r=\left(18.52^{+0.06}_{-0.06}\right)$~mag and $G_r=-0.25^{+0.10}_{-0.10}$ (Figure~\ref{fig:phase_function}b).  Again assuming solar colors as above, we find an equivalent $V$-band absolute magnitude of $H_V=(18.69\pm0.06)$~mag, making the effective radius of 457P's nucleus slightly larger than estimated by \citet{kim2022_p2020o1}, or $r_n\sim(0.56\pm0.06)$~km, assuming a $V$-band geometric albedo of $p_V=0.05$.

As this phase function fit is based on just four photometric points, additional measurements of 457P when it is inactive again to further refine its phase function parameters are highly encouraged.
However, given that this result still represents an advance over the calculations by \citet{kim2022_p2020o1} based on just one photometric point, we adopt the $H_V$ value from this phase function fit for all subsequent analyses presented in this manuscript.

\subsubsection{NIRCam Photometry}\label{section:nircam_photometry}

Photometric measurements of the first two of our four NIRCam observations of 457P that were visually determined to be relatively uncontaminated by background sources using $\rho=0\farcs3$ apertures result in measured average AB magnitudes of $m_{\rm F200W}=(23.40\pm0.08)$~mag and $m_{\rm F277W}=(24.03\pm0.01)$~mag, where the listed uncertainties correspond to the standard deviation of the individual photometric measurements included in each average.  These results yield a ${\rm F200W}-{\rm F277W}$ color of $-0.63\pm0.08$, similar to equivalent colors measured for 133P \citep[$-0.51\pm0.18$ and $-0.53\pm0.13$ during two separate visits;][]{hsieh2025_jwst133p}, and bluer than the equivalent colors measured for 238P \citep[$-0.38\pm0.05$;][]{kelley2023_jwst238p} and 358P \citep[$-0.29\pm0.05$;][]{hsieh2025_358p}, where this dichotomy appears plausibly attributable to the more substantial dust comae and larger water production rates (where the F277W bandpass includes the 2.7~$\mu$m water emission band) of 238P and 358P at the time of their observations compared to 133P and 457P.

\subsubsection{$Af\rho$ Calculations}\label{section:afrho}

In order to quantitatively characterize 457P's dust production rate evolution during its 2024 active apparition and use it to place our derived water production rate limit (see Section~\ref{section:volatile_species}) in context, we compute the quantity $A(0^{\circ})f\rho$, hereafter $Af\rho$, for each set of our optical observations.  $Af\rho$ is commonly used as a proxy for dust production rate that can be used to compare dust production activity levels derived from nucleus-subtracted and phase-function-corrected photometric measurements made of cometary coma observed at different times and under different conditions \citep{ahearn1984_bowell}.  It is given by

\begin{equation}
    Af\rho = {(2r_h\Delta)^2\over\rho} 10^{0.4[m_{\odot}-m_d(r_h,\Delta,0)]}
\label{eqn:afrho}
\end{equation}
where $r_h$ is in au, $\Delta$ is in cm, $\rho$ is the physical radius in cm of the photometry aperture used to measure the magnitude of the comet at the distance of the comet, $m_{\odot}$ is the apparent magnitude of the Sun at $\Delta=1$~au in the same filter used to observe the comet \citep[where we use $m_{\odot,V}=-26.71\pm0.03$;][]{hardorp1980_sun3}, and $m_d(r_h,\Delta,0)$ is the phase-angle-corrected (to $\alpha=0^{\circ}$) apparent magnitude of the dust with the flux contribution of the nucleus subtracted from the measured total magnitude.

$Af\rho$ is nominally independent of $\rho$ for a spherically symmetric, steady-state coma with a line-of-sight column density that scales with $\rho^{-1}$ (which results from a three-dimensional density profile that scales with $\rho^{-2}$) and no production or destruction of dust grains in the coma.  As described in \citet{hsieh2025_358p} and \citet{hsieh2025_jwst133p}, however, the theoretical aperture size-independence of $Af\rho$ is not preserved in the regime where coma morphology becomes dominated by radiation pressure effects \citep{jewitt1987_cometsbps,fink2012_afrho}, which in practice, applies to all of our ground-based observations. Instead, $Af\rho$ is expected to decrease with increasing $\rho$ at the image scales of our observations.

Given this consideration, in order to facilitate comparisons to previously published $Af\rho$ measurements, we follow the method used by \citet{hsieh2025_358p} of computing $Af\rho$ using a photometry aperture judged to be ideal for the particular data being measured (which ranged from $0\farcs7$ to $3\farcs2$; listed in Table~\ref{table:ground_photometry_457p_active}) for the data reported here; see Section~\ref{section:optical_data_reduction}), and extrapolating that $Af\rho$ measurement to a physical aperture size of $\rho=5000$~km at the distance of the comet (i.e., roughly comparable to aperture sizes typically used for published $Af\rho$ measurements; equivalent to $2\farcs7$ to $5\farcs1$ for the data we report here; listed in Table~\ref{table:ground_photometry_457p_active}).

\begin{figure}[ht]
    \centering
    \includegraphics[width=1.0\linewidth]{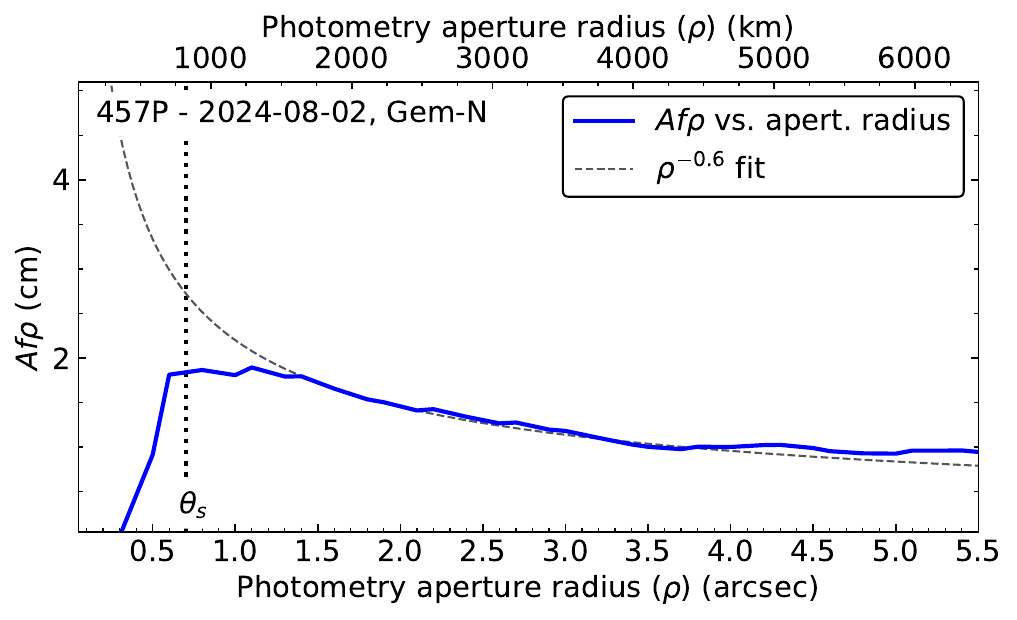}
    \caption{Plot of $Af\rho$ as a function of photometry aperture radius (solid blue line) in terms of arcseconds projected on the sky (bottom $x$-axis labels) and km at the distance of the comet (top $x$-axis labels) as measured for a composite image of 133P constructed from data obtained on UT 2024 October 27 by Gemini South, with a $\rho^{-0.6}$ power law (dashed black line) shown for reference.  The seeing ($\theta_s$) from Table~\ref{table:ground_observations_457p} is marked by a dotted vertical line.}
    \label{fig:afrho_profile}
\end{figure}

For this extrapolation, we examine a representative $Af\rho$ profile plotted as a function of $\rho$ that we measure for a relatively well-isolated detection (see Figure~\ref{fig:457p_optical_images}l) of 457P obtained on UT 2024 August 2 by Gemini North.  We find a good fit to the measured profile using a power law that scales with $\rho^{-0.6}$  (Figure~\ref{fig:afrho_profile}), and thus adopt this as the assumed power law form for the $Af\rho$ profiles for each of our observations (given the difficulty of conducting independent fits for data obtained in crowded fields).

\begin{figure*}[ht]
    \centering
    \includegraphics[width=0.65\linewidth]{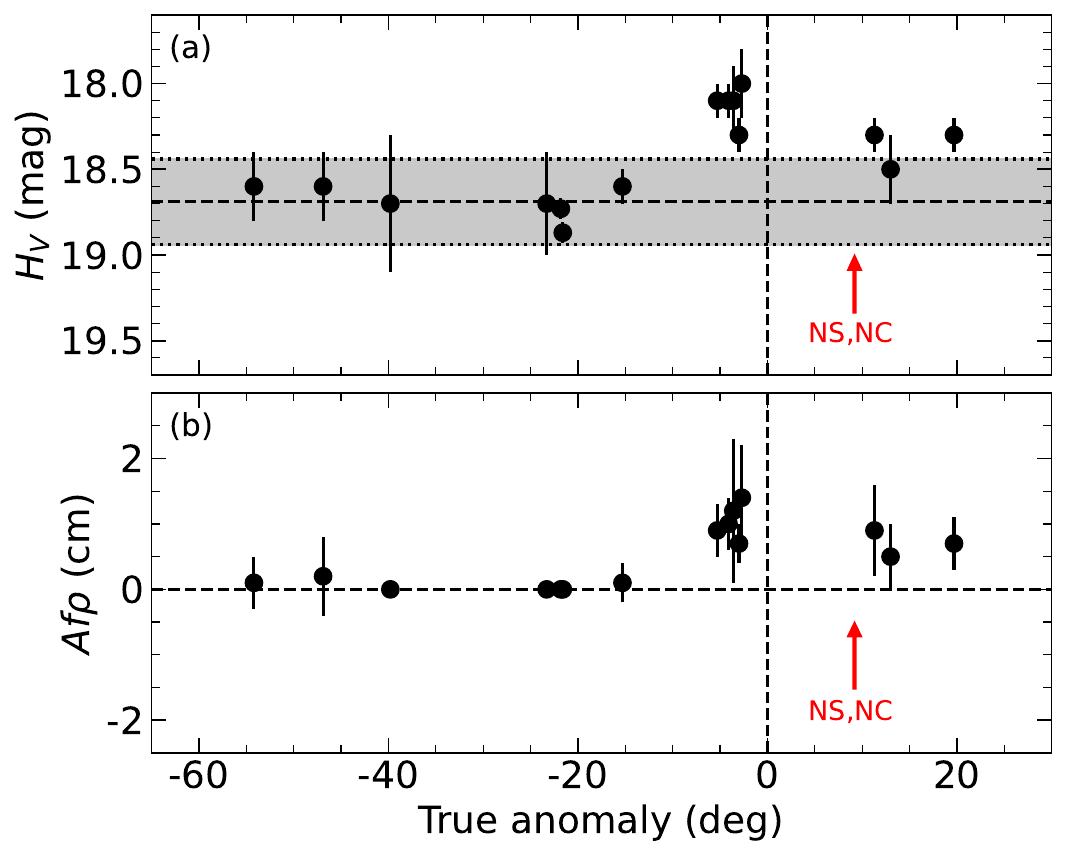}
    \caption{Plots of (a) equivalent total absolute $V$-band magnitude and (b) $Af\rho$ measured for 457P as functions of true anomaly.  In panel (a), the dashed horizontal line shows the absolute $V$-band magnitude of 457P's inactive nucleus of $H_V=18.69$~mag determined in this work (Section~\ref{section:nucleus_phase_function}), and the gray-shaded area bounded by horizontal dotted lines indicates the range of potential brightness variations allowed by the assumption of $\Delta m=0.25$~mag. In panel (b), the dashed horizontal line shows $Af\rho=0$, corresponding to the absence of activity.  In both panels, the true anomalies of our NIRSpec (NS) and NIRCam (NC) observations are marked.}
    \label{fig:absmag_afrho}
\end{figure*}

Following \citet{hsieh2025_358p}, to facilitate comparisons with other published work, we apply uncertainties for our $Af\rho$ values corresponding to aperture size ranges of $\rho=(5000\pm2000)$~km and add these in quadrature to the uncertainties arising from the underlying photometry.

These calculations were conducted for all sets of observations for which the measured flux-averaged total brightness of 457P was brighter than the expected brightness of its inactive nucleus at the midpoint of its rotational lightcurve based on the absolute magnitude of $H_V=18.69\pm0.06$ determined in this work (Section~\ref{section:nucleus_phase_function}).  For other measurements where the measured flux-averaged total brightness of 457P was fainter than the expected brightness of its inactive nucleus, we report $Af\rho$ values of 0~cm.  These $Af\rho$ values and the $H_V$ values upon which they are based are plotted as a function of true anomaly in Figure~\ref{fig:absmag_afrho}.
From Figure~\ref{fig:absmag_afrho}, we see that 457P appears to have become active sometime around or after $\nu\sim-20^{\circ}$ (approximately mid-June 2024 at $r_h\sim2.35$~au) and was clearly active by the time it reached $\nu\sim-5^{\circ}$ (in early August 2024 at $r_h\sim2.33$~au).
For comparison, detectable activity was determined to have started for 259P at $\nu\sim-45^{\circ}$ and $r_h\sim1.96$~au \rev{\citep[][see Section~\ref{sec:457P_in_context}]{hsieh2021_259p}}.

For reference, we also compute $Af\rho$ from our NIRCam observations obtained during the same visit as our NIRSpec observations.  In optical data from UT 2024 September 28 (our closest ground-base observations to the \jwst{} observations), the apparent $V$-band brightness of the dust comprised $\sim0.5$ of the total measured $V$-band brightness of the comet (due to the different assumed phase function behavior of the nucleus and the dust discussed in Section~\ref{section:optical_data_reduction}, this fraction differs from the one implied by the $A_d/A_n$ value listed in Table~\ref{table:ground_photometry_457p_active}, which was computed from absolute magnitudes).  Assuming the same contribution from the dust to the total brightness of the comet at near-infrared wavelengths, from the total measured magnitudes of the comet of $m_{\rm F200W}\sim23.40$~mag and $m_{\rm F277W}\sim24.03$~mag (Section~\ref{section:nircam_photometry}), we obtain apparent dust magnitudes of $m_{d,{\rm F200W}}\sim24.12$~mag and $m_{d,{\rm F277W}}\sim24.75$~mag, and phase-angle-corrected dust magnitudes of $m_{d,{\rm F200W}}(0^{\circ})\sim23.25$~mag and $m_{d,{\rm F277W}}(0^{\circ})\sim23.88$~mag, assuming the Schleicher-Marcus phase function (see Section~\ref{section:optical_data_reduction}).
Using Equation~\ref{eqn:afrho}, $m_{\odot}=-26.64$~mag for F200W, and $m_{\odot}=-26.03$~mag for F277W, we then compute $Af\rho_{\rm F200W}\sim5.4$~cm and $Af\rho_{\rm F277W}\sim5.3$~cm.  Using the same scaling factor as used by \citet{kelley2023_jwst238p}, these results correspond to $Af\rho\sim4.1$~cm at 0.7~$\mu$m.  Then extrapolating to $\rho=5000$~km assuming a $\rho^{-0.6}$ power law (see above), we finally find $Af\rho_{\rm 5000km}\sim1.0$~cm, which is in good agreement with results from ground-based optical data obtained around the same time ($Af\rho\sim0.9$~cm on UT 2024 September 28; Table~\ref{table:ground_photometry_457p_active}).

\subsubsection{Dust Production Rate Characterization}\label{section:dust_production}

We can use our photometric results to explicitly estimate the dust production rate following the methodology used by \cite{kim2022_p2020o1}. 
Re-writing Equation~\ref{eqn:afrho}, we first compute the absolute magnitude of the dust coma corresponding to our estimated $Af\rho$ value using a 5000~km-radius aperture at the time of the NIRSpec observations using
\begin{equation}
    m_{d,V}(1,1,0)=m_{\odot} - 2.5\log{(Af\rho)\rho\over (2r_h\Delta_{\rm cm})^2} - 5\log(r_h\Delta_{\rm au})
\end{equation}
where $r_h$ is in au, $\Delta_{\rm cm}$ and $\Delta_{\rm au}$ are in cm and au, respectively, and $\rho$ is in cm.  Using $Af\rho=(0.9\pm0.7)$~cm (Section~\ref{section:gas_to_dust}), we find $m_{d,V}(1,1,0)=19.0\pm0.8$.

We can then calculate the total cross section of the dust (in km$^2$) using:
\begin{equation}
    C_d = \frac{2.24 \times 10 ^{16} \pi}{p_V} 10^{0.4[m_{\odot,V}-m_{d,V}(1,1,0)]}
\label{eqn:cross_section}
\end{equation}
\citep{kim2022_p2020o1}, where we assume a geometric albedo of $p_V=0.05$, and use $m_{\odot,V}=-26.71\pm0.03$ \citep{hardorp1980_sun3}. Using this method, we find a total dust scattering cross section of $C_d=(0.7\pm0.6)$~km$^2$.

The dust mass, $M_d$, corresponding to this scattering cross-section can be estimated using
\begin{equation}
    M_d={4\over3}C_d{\bar a_d}\rho_d
\end{equation}
where ${\bar a_d}$ is the effective mean dust grain radius and $\rho_d$ is the dust grain bulk density, assuming that the coma is optically thin.  Assuming ${\bar a_d}=100$~$\mu$m and $\rho_d= 2500$~kg~m$^{-3}$ similar to that of CI and CM carbonaceous chondrites \citep{britt2002_astdensities_ast3}, we find a dust coma mass of $M_d=(2.4\pm1.8)\times10^{5}$~kg.

The dust velocity measured by HST during the 2020 apparition was found to be between $v_d=0.1$~m~s$^{-1}$ and $v_d=0.7$~m~s$^{-1}$ for particles with radii between $a_d=100$~$\mu$m and $a_d=1$~mm. Using $v_d=0.7$~m~s$^{-1}$ for $a_d=100$~$\mu$m dust grains, we can expect them to cross our aperture within $\sim7\times10^6$~s, which would require a $Q_d = (0.035\pm0.025)$~kg~s$^{-1}$, a factor of $\sim20$ smaller than 358P's dust production rate \citep{hsieh2025_358p}.  Using the same scaling for water vapor production, assuming a similar dust to gas ratio for 457P as for 358P, we would expect $Q_{\rm H_2O}\sim2.5\times10^{24}$~molecules~s$^{-1}$ for 457P, which is in remarkable agreement with the peak H$_2$O production rate of $Q_{\rm H_2O}\sim3\times10^{24}$~molecules~s$^{-1}$ (equivalent to $Q_{\rm H_2O}\sim0.09$~kg~s$^{-1}$, made assuming a dust-to-gas ratio of 10), estimated from HST observations in 2020 \citep{kim2022_p2020o1}, but is slightly larger than our calculated 3$\sigma$ upper limit of $Q_{\rm H_2O}<2.0\times10^{24}$~molecules~s$^{-1}$. Our observations achieved the desired sensitivity, but any H$_2$O activity must have been below this limit. \rev{This is somewhat surprising, given that the observations were made nearer to perihelion, and may be evidence that the availability of H$_2$O ice has diminished compared to the 2020 apparition. Whether that is due to a single decreased reservoir or a global retreat of the subsurface ice layer is difficult to assess from these snapshot observations. }

\citet{kim2022_p2020o1} found a peak dust production rate of $Q_d\sim0.9$~kg~s$^{-1}$ from HST imaging of 457P's dust comae during its 2020 active apparition, and determined that this dust production rate could be accounted for by an active area as small as $A_{\rm act}\sim1580$~m$^{2}$ in size, assuming a dust-to-gas ratio of $f_{dg}=10$ \citep[e.g., see][]{jewitt2014_133p}. Our corresponding dust production rate for the JWST observations, $Q_d\sim0.035$~kg~s$^{-1}$, would require an active area of $A_{\rm act}\sim1200$~m$^{2}$ if we assume that the dust-to-gas ratio is more similar to that of 358P \citep[$f_{dg}\sim0.5$;][]{hsieh2025_358p}, but the upper limit H$_2$O production rate we find for 457P ($Q_{\rm H_2O}<0.06$~kg~s$^{-1}$; Section~\ref{section:volatile_species}) implies an active area of $A_{\rm act}<5130$~m$^{2}$, assuming sublimation from a pure H$_2$O ice surface in equilibrium with sunlight in the \citet{cowan1979_cometsublimation} framework\footnote{https://ice-sublimation-tool.astro.umd.edu/}. A dust-to-gas ratio greater than $f_{dg}\sim0.7$ is required to reconcile our non-detection of H$_2$O with the observed NIRSpec dust morphology, which indicates slow (and therefore, likely large) grains, similar to the results of the 2020 HST analysis.

Due to the observing geometry, our synchrone/syndyne analysis using the {\tt sbpy} syndyne package shows substantial degeneracies between $\beta$ and duration of activity that satisfy the observed morphology in the NIRCam dataset. However, the narrow tail width is indicative of very low dust ejection speeds corresponding to dust grains with radii between $a_d=0.01$~mm and $a_d=1$~mm, consistent with the grain sizes reported by \citet{kim2022_p2020o1}. As shown in Figure~\ref{fig:457p_optical_images} and Table \ref{table:ground_photometry_457p_active}, 457P was already clearly active by UT 2025 August 2, 
meaning that the onset of activity was at least $\sim50$~days prior to our \jwst{} observations, and that
the tail is not due to the onset of activity less than two days prior to our \jwst{} observations, as shown in the synchrone analysis (Fig. \ref{fig:grain_size_syndyne_synchrone}).

\subsection{Morphology}\label{section:morphology}

\begin{figure*}[ht]
    \centering
    \includegraphics[width=0.75\linewidth]{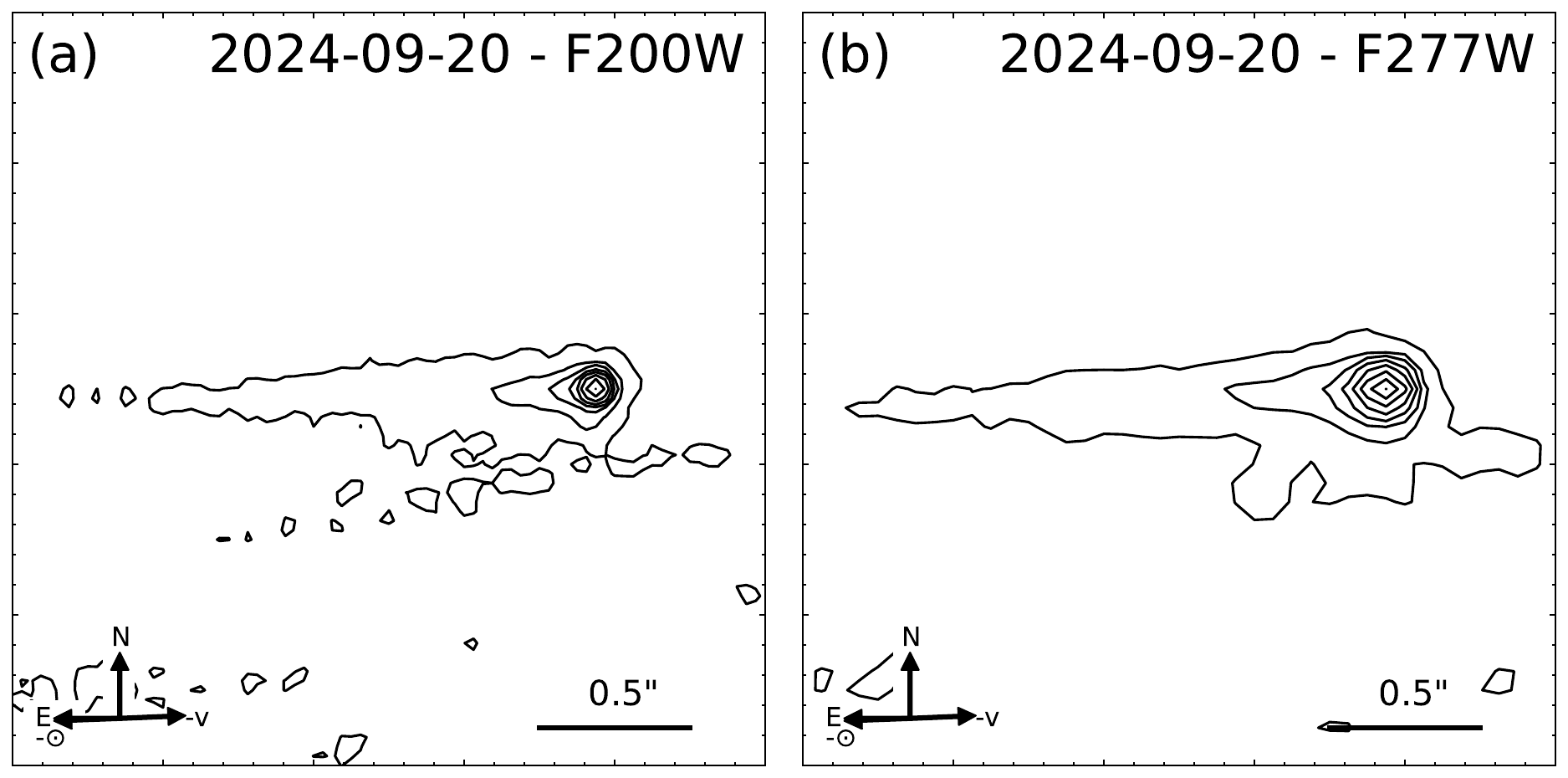}
    \caption{Contour plots (using 8 logarithmically spaced contour levels ranging from the peak value of each image --- 5.9~MJy/sr for F200W and 1.4~MJy/sr for F277W --- to the background level --- $\sim0.7$~MJy/sr for F200W and $\sim0.4$~MJy/sr for F277W --- for each image) of the inner coma of 457P constructed from (a) F200W and (b) F277W NIRCam median composite images obtained on UT 2024 September 20, shown in Figures~\ref{fig:nircam_images}a and \ref{fig:nircam_images}b, respectively.  A $0\farcs5$ angular scale bar (790~km at the distance of the comet) is shown in each panel, where the orientations of the images and contour plots are the same.}
    \label{fig:nircam_contours}
\end{figure*}

Images (Figure~\ref{fig:nircam_images}) and contour plots (Figure~\ref{fig:nircam_contours}) of the inner coma of 457P as imaged by NIRCam show that the comet's surface brightness profile deviates from circular symmetry beyond $\rho\sim0\farcs1$ from the photocenter in both the F200W and F277W median composite images.  Beyond this distance from the photocenter, a visible but low-brightness tail extends at least $\sim2''$ towards the East (i.e., a position angle of ${\rm PA}\sim90^{\circ}$ East of North), or essentially aligned with the direction of the anti-Solar vector and in the opposite direction of the comet's negative heliocentric velocity vector (Table~\ref{table:jwst_457p_observations}; Figure~\ref{fig:nircam_images}).

Given the tight upper limits on volatile production rates, the distinct anti-sunward tail present in the NIRCam images is intriguing. The narrow appearance and near-complete alignment with the anti-solar vector suggest low initial velocity dust that is swept away by radiation pressure. Given the narrow morphology of the tail it becomes difficult to estimate the distribution of active areas and rotational axis of the nucleus, as is typically tested with a cometary dust dynamics model \citep{moreno2013_p2012t1,hsieh2025_358p}. 

\begin{figure}
    \centering
    \includegraphics[width=0.95\linewidth]{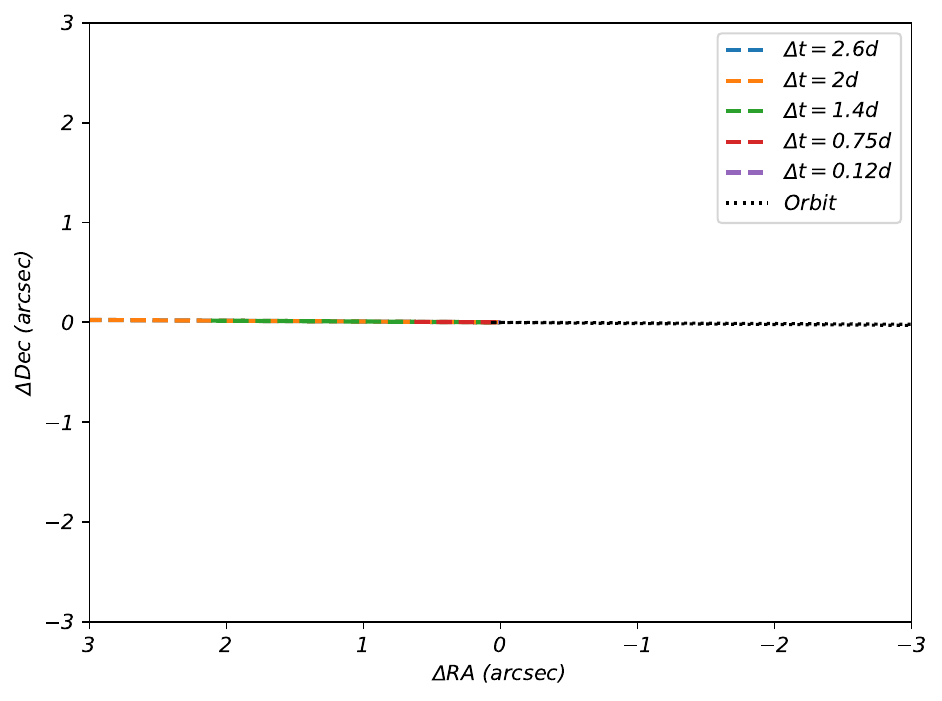}
    \includegraphics[width=0.95\linewidth]{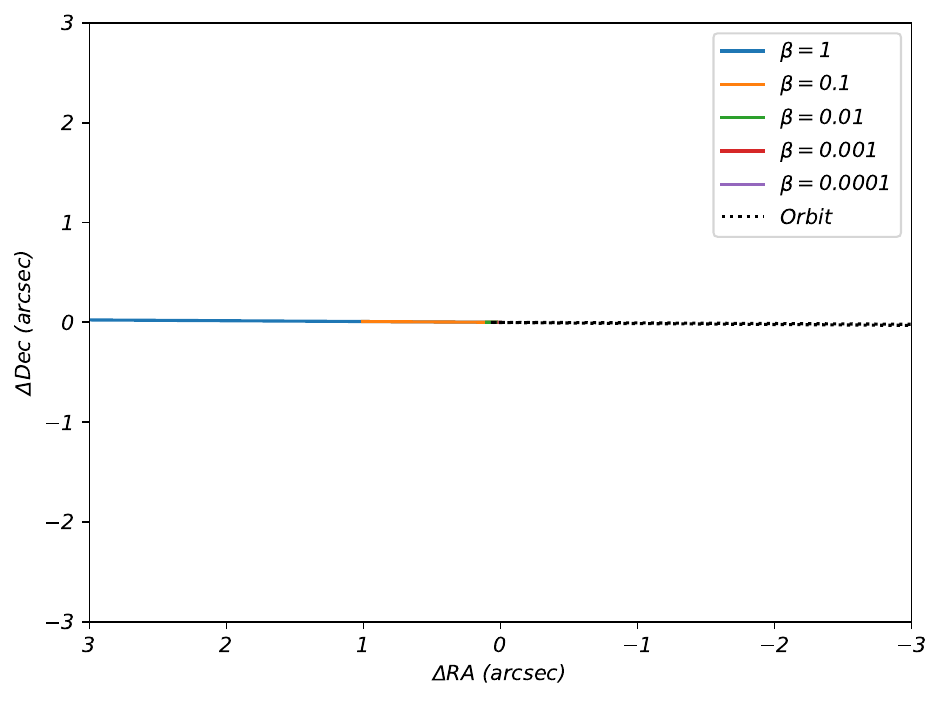}
    \caption{ Computed synchrones (top) and syndynes (bottom) for the observing geometry of our NIRCam observations with the FOV of our NIRSpec IFU observations (3\farcs$\times$3\farcs), detailed in Table \ref{table:jwst_457p_observations}, showing particles varying in $\beta$ released from the surface of 457P at the same time (synchrones) and particles released over time with set $\beta$ values (syndynes) using the {\tt sbpy} syndyne and synchrone tools. The observing geometry makes it difficult to rule out smaller $\beta$ values for our dust grains, and results in a somewhat comical plot, but the extension of the tail in Fig. \ref{fig:nircam_images} seems to be indicative of large ($\beta<$0.01) grains that are continuously released. Dust activity visible in Figure~\ref{fig:457p_optical_images} and indicated by photometric results listed in Table \ref{table:ground_photometry_457p_active} show activity reported prior to our JWST observations, indicating that recent onset of activity with smaller grains is unlikely. }
    \label{fig:grain_size_syndyne_synchrone}
\end{figure}

An additional source of model degeneracies lies within the dust size distribution. If these are small grains of dust being swept back quickly by the solar radiation pressure, as they experience the largest acceleration, then they must be recent escapees from the surface of 457P. However, we cannot exclude that these may be larger grains accelerated more slowly; the larger the grains, the more time that may have passed since the potential cessation of activity and our JWST observations. This scenario offers its own constraint, because it would require the peak of activity to occur quite close to perihelion before ending rather abruptly. We present a graphical summary of this in Figure \ref{fig:grain_size_syndyne_synchrone} to illustrate the relationship between last active date and minimum size grain remaining in the observed tail. Conversely, the lack of an extended tail in the anti-solar direction in Fig. \ref{fig:nircam_images} seems to confirm conclusions from the ground-based photometry that there was not more substantial activity of 457P in the days prior to our observations. 

\subsection{Gas to Dust Ratios}\label{section:gas_to_dust}

The relationship between gas and dust production rates, as parameterized by $Af\rho/Q_{\rm H_2O}$, have been computed for other MBCs \citep{kelley2023_jwst238p,hsieh2025_358p,hsieh2025_jwst133p}, and so it is useful to compute them for 457P as well for comparison.
Unfortunately, we were unable to acquire usable observational data closely bracketing our NIRSpec observations, although we did acquire usable photometry relatively close in time (UT 2024 September 28) from Gemini South, and as such, adopt the measured value of $Af\rho=(0.9\pm0.7)$~cm from those observations as the closest available characterization of the dust production rate at the time of our NIRSpec observations.
From this, we compute $\log_{10}(Af\rho/Q_{\rm H_2O})>-24.4\pm0.3$, a similar dust production with respect to water vapor production as previously measured MBCs \citep{kelley2023_jwst238p,hsieh2025_358p,hsieh2025_jwst133p}.

\section{Discussion}

\subsection{Implications of the Non-Detection of Water Vapor}\label{sec:nondetection_implications}

The non-detection of H$_2$O outgassing to the level of $Q_{\rm H_2O}<2\times10^{24}$~molecules~s$^{-1}$, as well as the non-detections of any other major cometary volatile species, is surprising given the observational confirmation in this work that 457P is recurrently active near perihelion \citep[considered to be a strong indicator of at least partially sublimation-driven activity; e.g.,][]{hsieh2012_scheila,jewitt2024_continuum_comets3}, and evidence of activity at the time of our non-detection of water vapor outgassing by NIRSpec in the form of a visible dust tail in NIRCam observations (Figure~\ref{fig:nircam_images}) and photometric excesses measured from ground-based observations (Section~\ref{section:afrho}).  Given these results, it is necessary to evaluate whether the non-detection of water vapor outgassing is consistent with the weak dust activity observed at the same time, and if not, what additional physical circumstances could help to explain the observed dust activity in the absence of detectable outgassing.

Explaining how such particles could be continuously produced by such apparently weak sublimation, if any, is a challenge. In the following sections, we consider the roles that jet-like dust emission and rapid rotation could play in helping to explain 457P's activity.

\subsection{Collimated Dust Ejection}\label{section:dust_jets}

One possible explanation for reconciling 457P's observed dust activity with the absence of detectable outgassing is jet-like gas emission imparting more force on dust particles than the same amount of gas being emitted isotropically. For this explanation to be tenable, the point at which the original ejection direction of the dust is fully erased by the action of radiation pressure would need to be unresolvable, even by \jwst{}, making the observed morphology indistinguishable from that expected from isotropic emission. This would essentially require the existence of a narrow jet that is responsible for the entire dust production, with a water production rate just below our sensitivity, with a dust-to-gas ratio larger than $f_{dg}\sim0.7$. Such a scenario is directly testable with our NIRCam images; if present, the dust particles produced by the jet must be accelerated tailward before they are capable of producing an asymmetric feature in the NIRCam images. From our NIRCam images, that distance appears to be $<500$ km.  By rearranging the turnback distance equation to solve for velocity, we can determine if the dust ejection velocity is achievable at the maximum allowable H$_2$O production rate. 

The turnback distance over which dust grains ejected sunward with terminal ejection velocity of $v_g$ are turned back by solar radiation pressure is given by
\begin{equation}
    X_R \sim { v_g^2 \over 2 \beta_d g_\odot} \cdot \left( r_h \over 1 \, {\rm au} \right)^2
    \label{equation:turnback_dist}
\end{equation}
\citep[see][]{jewitt1987_cometsbps}, where $g_{\odot}=0.006$~m~s$^{-2}$ is the gravitational acceleration to the Sun at $r_h=1$~au,
$\beta_d$ is the ratio of a particle's acceleration due to solar radiation pressure to the acceleration due to the Sun's gravity \citep{burns1979_dustradiationforces}, and $v_g\propto\beta_d^{1/2}$ for dust particles that are accelerated by outflowing gas, meaning that $X_R$ as shown in Equation~\ref{equation:turnback_dist} is essentially independent of $\beta_d$.  Comet dust modeling typically implement $\beta_d$ as a proxy for particle sizes, where $a_d\approx\beta_d^{-1}$ for $a_d$ in $\mu$m. 

Solving Equation \ref{equation:turnback_dist} for $v_g$, we find

\begin{equation}
    v_g^2 \sim \sqrt{X_R \cdot 2 \beta_d g_\odot \left( r_h \over 1 \, {\rm au} \right)^{-2}}
    \label{equation:turnback_ejection_vel}
\end{equation}

\begin{figure}
    \centering
    \includegraphics[width=1.0\linewidth]{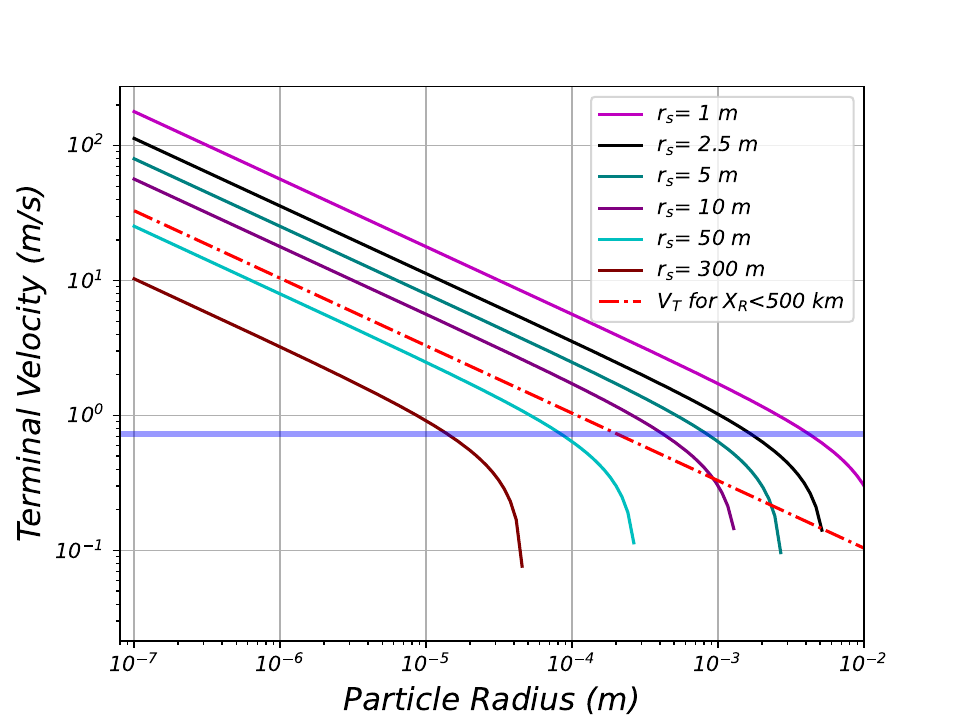}
    \caption{Terminal dust velocities for particles between $a_d=0.1$~$\mu$m and $a_d=1$~cm for various active area radii with a total $Q_{\rm H_2O} \leq0.06$~kg~s$^{-1}$, as calculated for 457P during our observations using the small-source approximation from \citet{jewitt2014_133p}. \rev{Note that this limitation on total production rate reduces the number density of H$_2$O for larger active areas, in turn reducing gas pressure and terminal velocity.} The maximum terminal velocity for a given particle size to have $X_R<500$~km is plotted as a dotted line. If dust production was driven by sublimation below our upper limit for H$_2$O, the dust must come from an active area with radius $r_s\geq50$~m unless dominated by grains with $a_d>500$~$\mu$m. The velocity of a particle immediately lost from the surface of 457P due to rapid rotation is plotted s a horizontal blue line at 0.73 m s$^{-1}$.  }
    \label{fig:vt_ssa}
\end{figure}

Using $X_R=500$~km, $g_{\odot}=0.006$~m~s$^{-2}$, and $r_h$ = 2.335 au,  we then find

\begin{equation}
    v_g(\beta) = 10.4\beta^{1/2}~\mathrm{m}~\mathrm{s}^{-1}
    \label{equation:v_g_max}
\end{equation}
We therefore find a maximum ejection velocity of $v_g\sim10$~m~s$^{-1}$ for a particle with a $\beta=1$ ($a_d\sim1$~$\mu$m) that will still produce a turnback distance of $X_R=500$~km. For the larger particles expected at 457P, based on the previous HST data obtained by \citet{kim2022_p2020o1}, of 100 $\mu$m, the maximum ejection velocity is $v_g\sim1$~m~s$^{-1}$. Assuming a circular active region with a radius of 5 m, our upper limit H$_2$O production rate could, in principle, accelerate particles to their maximum allowable velocity while still having a turnback distance of $X_R\leq$ 500 km, using the small source approximation detailed in Equation A5 of \citet{jewitt2014_133p}. We show several other active region sizes plotted against Equation \ref{equation:v_g_max} in Figure \ref{fig:vt_ssa}. For the relevant H$_2$O production rate, it is worth noting that at too large of an active area ($r_s \geq300$~m), there is no longer enough pressure to continuously loft any dust with $a_d\geq$10~$\mu$m\footnote{\rev{Note that our Figure 11 differs from \citet{jewitt2014_133p} by keeping the total production rate of the active area constant, rather than the production rate per square meter. For our case, limited by our upper limit on the water production, a larger active area means a lower number density of water molecules to accelerate the dust.}}. Intriguingly, the upper limit constraint placed on the active area by our upper limit on $Q_{\rm H_2O}$ (Section~\ref{section:volatile_species}) is quite close to 50~m ($r_2$ in Figure~\ref{fig:vt_ssa}), which is capable of accelerating dust particles with $a_d<0.1$~mm up to, but not past, the maximum velocity limit imposed by our turnback distance limit.

If we add the grain size determination from \citet{kim2022_p2020o1} into our priors, which shows that grains between $a_d=0.08$~mm and $a_d=1.14$~mm need to be lofted, we require a smaller active area, which in turn requires that smaller dust grains have a larger velocity, increasing the resulting turnback distance. There is no reasonable active area radius that can loft dust particles of the size range found by \citet{kim2022_p2020o1} without producing a more extended coma from smaller grains, which are not seen in the NIRCam data, unless they are not present in the surface regolith in the first place.

\subsection{Centrifugal Dust Ejection}\label{section:centrifugal_ejection}

In addition to, or possibly instead of, using collimated dust emission to explain the unexpectedly low gas production limit, we consider that the estimated rotation period of 457P, $P_{rot} \approx$1.67 hrs \citep{kim2022_p2020o1}, is rapid enough to lower the required force from sublimating volatiles to lift the same dust grains. On its own, the rapid rotation rate cannot explain the dust production; whatever produces the dust must only be effective near perihelion. However, given that the exact conditions for lofting dust grains are poorly understood \citep[c.f.][]{agarwal2023_dustemission_comets3}, we examine this factor analytically. The centrifugal acceleration, $a_{cent}$, experienced by a dust grain \rev{at the equator} of 457P can be expressed as
\begin{equation}
    a_{cent}=  \omega_{457P}^2r_{457P}
\label{eqn:centrifugal}
\end{equation}
where $\omega_{457P}$ is the angular velocity of 457P, and $r_{457P}$ is the nucleus radius (assuming a spherical body). For 457P's rotational period of $P_{rot}$=1.7 hrs and radius of $\sim$560 m, that would produce a centrifugal acceleration of 5.9$\times$10$^{-4}$ m s$^{-2}$.  Assuming a spherical shape and C-type asteroid density \citep{carry2012_astdensities} for 457P, that dust particle would experience a gravitational acceleration of 3.9$\times$10$^{-4}$ m s$^{-2}$, slightly lower than the centrifugal force. \rev{By solving for the latitude at which the centrifugal force is equal to the gravitational force, where $r_{457P}$ is substituted by $r_{457P}$cos($\phi$), where $\phi$ is the latitude, again in the simplifying assumption of a spherical body, we can determine where on the ejected particles would be released from.} This suggests that \textit{any} loose particles on the surface within 54$^{\circ}$ of the equator could be ejected without invoking any sublimation processes. This was first proposed by \citet{kim2022_p2020o1} upon their detection of a possible 1.67 hr rotational period, but takes on new significance in the light of our non-detection of H$_2$O during dust activity. The probable rotation rate of 457P is rapid enough to eject material, but that centrifugal force will not accelerate material once contact is lost with the surface; at best that provides an initial velocity for particles of 0.73 m s$^{-1}$, assuming ejection from the equator of 457P, below the velocity required for a turnback distance of 500 km for particles smaller than 20 $\mu$m. Given that much of the coma appears to be made of larger particles, based on the extended and condensed tail in the NIRCam images as well as the prior HST images, additional insolation-driven activity hypotheses are needed.

\subsection{Other Potential Contributing Factors}\label{section:other_factors}

One basic possibility that we have not yet considered is simply that 457P's gas production had indeed nearly ceased by the time of the NIRSpec observations, and the dust tail and coma observed at the time of those observations consisted of dust that had been ejected some time ago and had not yet had time to dissipate.  As computed in Section~\ref{section:dust_production}, particles with $a_d\sim100$~$\mu$m are expected to have coma crossing times of $\sim7\times10^6$~s, or $\sim$80~days, meaning that such particles could have been ejected more than two months prior to the \jwst{} observations at a time of higher water vapor production, and would still be present, providing the appearance of current dust ejection activity, during the \jwst{} observations when water vapor production may have been much weaker or even entirely absent.

Alternatively, thermal stresses near perihelion, perhaps induced by a high obliquity rotational axis producing a rapid change to the thermal environment, may be responsible for producing a reservoir of loose dust available for ejection by centrifugal force without any corresponding sublimation. Thermal fatigue was hypothesized as a source for the activity observed around Bennu by the OSIRIS-ReX mission, driven by the thermal cycling of surficial boulders \citep{molaro2020_bennuactivity}, and may be a relevant process even at the distances and temperatures of the main asteroid belt \citep{molaro2015_thermalfracture,jewitt2022_continuum_comets3}. Such a process could become more prominent at perihelion, but why it would occur for 457P at its perihelion distance of 2.3 au is unclear.

\subsection{Placing 457P/PANSTARRS in Context}\label{sec:457P_in_context}

The prevailing theories of the outer main-belt formation all involve implantation of carbonaceous chondrite (CC) -like objects during the giant planet instability \citep{gomes2005_nicemodel,walsh2011_grandtack,izidoro2016_asteroidbeltchaos}. The CC group likely formed in a spatially separated reservoir from the non-carbonaceous (NC) chondrites based on isotopic ratios \citep[e.g.][]{kleine2020_NCCCmeteorites,nanne2019_NCCCdichotomy,worsham2019_CCNCreservoirs}. Given the difficulty of implanting contemporary JFCs into circular orbits in the outer main-belt to explain the presence of MBC activity \citep{hsieh2016_tisserand}, it follows that some MBCs must represent an earlier implanted or natal population. For all MBCs observed by \jwst{} thus far (238P, 358P, and 457P), constraints on H$_{2}$O production rates all indicate a substantial level of depletion relative to JFCs. 

Among the most well-studied MBCs, 259P/Garradd ($a=2.727$~au, $q=1.796$~au, $e=0.342$, $i=15\fdg899$, $T_J=3.217$) is the most dynamically similar to 457P, where 259P was the first MBC to be identified with a semimajor axis interior to the 5A:2J mean-motion resonance with Jupiter at 2.824~au \citep{jewitt2009_259p}, and the first MBC in this region confirmed to exhibit recurrent activity near perihelion \citep{hsieh2021_259p}.  A detailed observational study of 259P from approximately four months prior to its 2017 perihelion passage to five months after that perihelion passage (covering a true anomaly range of $-55\fdg6<\nu<59\fdg7$) by \citet{hsieh2021_259p} found a likely activity onset point $\sim104$ days prior to perihelion when the object was at $r_h=1.96\mp0.03$~au and $\nu=313\fdg9\pm0\fdg4$, and an estimated dust mass loss rate of ${\dot M_d}=(4.6\pm0.2)$~kg~s$^{-1}$.  In that work, the authors noted that the heliocentric distance of 259P's activity onset point ($r_h=1.96$~au) was significantly smaller than those of other MBCs ($r_h\sim2.5-2.6$~au), and suggested that that could indicate that its ice reservoirs could be located at greater depths than on MBCs farther from the Sun, increasing the time required for a Solar-irradiation-driven thermal wave to reach those reservoirs and initiate sublimation.  This deeper ice, in turn, could be due to more rapid ice depletion caused by the object's closer proximity to the Sun, and therefore higher average temperatures around its orbit, relative to other MBCs.

For comparison, 457P shows a more complicated story. If present, H$_2$O sublimation must fulfill a very strict set of criteria to explain the observed dust production rate (small active area, capable of producing large grains), that still occupies a relatively large parameter space when considering the possible rapid rotation rate of the object. While the possible 1.67 hour rotation period would make it easier to liberate material from the surface, the mechanism to waste the surface away and provide a continuous source of dust near perihelion is not well constrained; rotation rates that high would indicate a high cohesive strength to the nucleus to remain intact without any global failures \citep{zhang2022_bennu_shape}. If a deep ice reservoir is to be invoked to explain 457P's activity, it must be exiting the surface at a very localized region to explain the large grains in the dust tail. 

\begin{figure}
    \centering
    \includegraphics[width=0.99\linewidth]{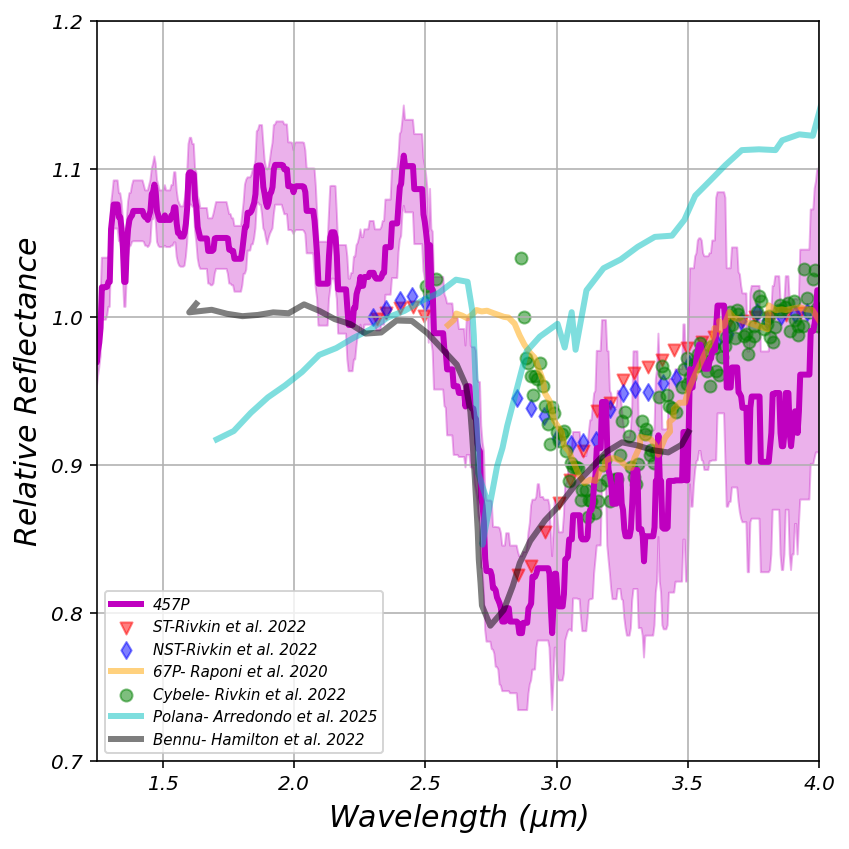}
    \caption{\rev{Relative reflectance spectra for 457P/Lemmon-PANSTARRS, the average sharp-type (ST) non-sharp type (NST) spectra, Cybele \citep{rivkin2022_3micron}, 67P/Churyumov-Gerasimenko \citep{raponi2020_cometorganics}, (142) Polana \citep{arredondo2025_jwst_polana}, and (101955) Bennu from OVIRS \citep{Hamilton2022}. We note that the redward edge of the 3 $\mu$m contains complex absorption features, similar to Cybele, 67P, and the NST spectra, while the 2.7 $\mu$m region is notably broader than (142) Polana, indicating strong hydration absorption with NH-bonds absorbing nearer 3.1 $\mu$m. We also plot a global reflectance spectrum of Bennu \citep{Hamilton2022} to highlight overall similar band shape, but a substantially different band center.}}
    \label{fig:ref_comp_overview}
\end{figure}

The apparent paucity of small dust around 457P, the lack of volatiles, and some similarity of the 3 $\mu$m band to the nonsharp-type (NST) spectra defined by \citet{rivkin2022_3micron}, rather than the shape of the ST absorption \citep{arredondo2024_polana_henrietta,arredondo2025_jwst_polana}, may indicate a shared formation or evolution between 457P, 238P, and 358P. All prior observations of 457P support sublimation as its activity driver \citep{weryk2020_p2020o1,kim2022_p2020o1}, and yet our observations place tight limits on the water production rate and distribution that are an order of magnitude lower than H$_2$O production rates measured for other MBCs. \rev{We note that while the average ST spectrum has some resemblance to the 2.7-3.1 $\mu$m region for 457P, comparing directly to JWST NIRSpec observations of ST (142) Polana shows key differences in band center, width, and shape (Fig. \ref{fig:ref_comp_overview}. There also appears to be similarity in the band shape of (101955) Bennu as observed in situ with OVIRS on board OSIRIS-REx \citep{Hamilton2022}, but not in the band center (2.74 $\mu$m for Bennu, compared to 2.88 $\mu$m for 457P), likely due to differences in the OH-bearing phyllosilicates. This is not terribly surprising, given Bennu's outer main belt origin and dynamical evolution into the inner solar system, but interesting to note for future studies of main belt objects with JWST given Bennu's likely different collisional and thermal evolution \citep{bottke2015_bennu,Walsh2024}, and the possiblity that Bennu may be a space-weathered MBC analog. } \citet{rivkin2022_3micron} hypothesized that the discrepancy between the ST and NST near-infrared reflectance spectra of main-belt asteroids was indicative of formation conditions, with ST and NST objects forming inside and outside the ammonia ice line, respectively, producing fundamentally different structure in their 3 $\mu$m absorption bands. In 457P, we have found an active MBC nucleus that resembles these NST spectra and that appears similar to the comet 67P/Churyumov-Gerasimenko \citep{raponi2020_cometorganics}.  However, we note that the band shape is not well matched by any of the JWST spectra currently published for main-belt asteroids \citep{arredondo2024_polana_henrietta,rivkin2025_jwst_mbas,arredondo2025_jwst_polana}\rev{, but bears some band shape, if not band center, similarities to Bennu.} If we follow that 457P's complex absorption band is most similar to the the C-type complex \citep{rivkin2022_3micron} and that those have been proposed to be the progenitors of most CM meteorites, \citep{vernazza2016_cmparentbodies,vernazza2017_ctypeasteroids}, a small logical leap is required to the conclude that we may already have samples of previously active main-belt comets in the meteoritic collection. \rev{Ultimately testing these theories will require more high sensitivity observations of the outer main belt, both active and inactive. Further NIR surveys of the MBC nuclei to clarify band properties of the 3 micron region, as well as more targeted modeling for 457P, are required before the mineralogy can be accurately assessed, but these spectra suggest a complex mixture of phyllosillicates and organics. }

457P's repeated activity near perihelion for \rev{three} apparitions is strong evidence that sublimation is the likely source of activity, even if gas was not directly detected. Occam's razor suggests that the liberation of dust when thermal conditions are warmest is indicative of sublimation, and the likely rapid rotation of 457P works in favor of ejecting dust by lowering the required gas drag. Notably, there is evidence for repeated activity from a non-sublimating asteroid, (6478) Gault \citep{chandler2019_gault}. However, we can draw a clear distinction between Gault and 457P, as the repeated activity from Gault was not correlated with its perihelion passages. 

When placed in the context of previous \jwst{} observations of 238P and 358P, both with clear detections of water vapor and relatively large dust comae, it becomes clear that more observations of the MBC nuclei, as well as comae, are needed to make progress in identifying the source regions of the MBCs. We can propose the following hypothesis: the activity of 238P and 358P, which we know is due to sublimation of H$_2$O, is due to those objects having compositions consistent with formation beyond the ammonia ice line, and therefore any observations of their nuclei with \jwst{} should reveal a 3 $\mu$m absorption feature more consistent with the NST observed on 67P, Themis, or Ceres, as seen with 457P. However, if 238P and 358P also show different 3 $\mu$m absorption bands indicating formation in a region depleted in ammonia, then we cannot conclude that the differences in activity between the current MBC sample observed with \jwst{} are due to thermal evolution alone and must invoke different formation regions for MBCs. \rev{The dust surrounding these objects can mask this absorption feature, as cometary dust will contribute a red, linear reflectance to the extracted spectrum; once the total surface area of the dust in the aperture becomes of the same order as the surface area of th nucleus the effect becomes significant. Modeling this dust contribution to remove it effectively requires a Mie scattering model \citep[see e.g.][]{protopapa2018_C2013US10, kareta2023_46pcoma}. Future JWST observations of these objects at aphelion, where their dust coma should be dissipated, would be able to avoid this extra correction and reduce uncertainty in the derived band properties. }

\section{Summary}

In this work, we present the following key findings from our ground-based and JWST observing campaign:
\begin{enumerate}
\item{We report 3$\sigma$ upper limits of four common cometary volatiles from our NIRSpec spectrum: $Q_{\rm H_2O}<2.0\times10^{24}$~molecules~s$^{-1}$; $Q_{\rm CO}<4.8\times10^{24}$~molecules~s$^{-1}$; $Q_{\rm CO_2}<1.4\times10^{23}$~molecules~s$^{-1}$, and $Q_{\rm CH_3OH} <2.2\times10^{24}$~molecules~s$^{-1}$.}
\item{Ground-based photometry shows weak activity from 457P during the 2024 perihelion passage, with $Af\rho$ between 0.1 and 1.5 cm for $-$6$^{\circ} < \nu  < 19.7^{\circ}$. Ground-based photometry during periods of inactivity allowed us to derive an updated $H_V=18.69\pm0.06$, corresponding to an equivalent circular radius of $r_{\rm 457P}=(0.56\pm0.06)$~km, assuming a geometric albedo of $p_V=0.05$. }
\item{NIRCam F200W and F277W images show a short ($<5''$) narrow dust trail at the time of our NIRSpec observations. Syndyne, synchrone, and active region analyses in the context of the continuous weak dust activity observed by the ground-based campaign, as well as previous observations from \citet{kim2022_p2020o1}, suggest that this trail is likely comprised of large dust grains with $\beta < 0.01$. }
\item{Observations of a weak dust coma and our updated nucleus radius derivation are consistent with the match we find between the modeled and observed reflectance spectrum, with $Af\rho < 1$ cm expected to contribute less than 2$\%$ to the measured spectrum.}
\item{We identify an asymmetrical 3 $\mu$m absorption feature with a band center at 3 $\mu$m and a bandwidth of $\sim$1.5 $\mu$m, with an absorption depth of $\sim$20$\%$ relative to the continuum. This absorption feature is similar to those seen on 67P, Ceres, Themis and other non-sharp type objects \citep{raponi2020_cometorganics,rivkin2022_3micron}. }

\end{enumerate}

The paradigm that asteroids and comets occupy a continuum continues to be supported by new evidence, and here we highlight that while spectrally 457P appears to be similar to the low-albedo nonsharp-type asteroids identified by \citet{rivkin2022_3micron} as well as 67P \citep{raponi2020_cometorganics}, its activity during the 2024 perihelion passage, while weak, may have been driven by H$_2$O sublimation below \jwst{}'s sensitivity. However, if true, the sublimation must have been from an extremely localized area to loft grains of the right size, even when aided by the centrifugal force of 457P's likely rapid rotation. Those specific conditions for cometary activity from objects that look otherwise asteroidal, and cometary in NIR reflectance, are a major motivation to increase the sample size of known MBCs with surface reflectance data. We propose that the more active 238P and 358P, with their outer main-belt origins, will have surface reflectance properties more similar to NST objects like Cybele and 67P/Churyumov-Gerasimenko. Testing this hypothesis with \jwst{} is a high priority for evaluating the location and abundances of remaining volatile reservoirs in the main-belt.

\section*{Data and Software Availability\label{section:software}}
Pipeline-processed \jwst{} data are publicly available from the Space Telescope Science Institute's Mikulski Archive for Space Telescopes at \url{https://mast.stsci.edu/} under JWST program ID 5551, and at \dataset[https://doi.org/10.17909/t50v-v376]{https://doi.org/10.17909/t50v-v376}.
Ground-based image data from Gemini Observatory are publicly available online from the Gemini Observatory Archive at \url{https://archive.gemini.edu/} under program IDs GS-2023A-LP-104, GS-2024A-Q-111, GN-2024B-Q-114, and GS-2024B-Q-113.

This work makes use of the Planetary Spectrum Generator at \url{https://psg.gsfc.dnasa.gov/}, the Ice Sublimation Model at \url{https://github.com/Small-Bodies-Node/ice-sublimation}, and the \jwst{} science data calibration pipeline at \url{https://github.com/spacetelescope/jwst/}.
This research also makes use of the Jet Propulsion Laboratory's Horizons online ephemeris generation tool \citep{giorgini1996_horizons};
NASA's Astrophysics Data System Bibliographic Services\footnote{\url{https://ui.adsabs.harvard.edu/}}, which is funded by NASA under Cooperative Agreement 80NSSC21M00561;
{\tt astropy}, a community-developed core {\tt python} package for astronomy; {\tt ccdproc}, an {\tt astropy} package for image reduction;
{\tt L.A.Cosmic}, a cosmic ray rejection algorithm \citep{vandokkum2001_lacosmic};
{\tt pyraf}, a product of the Space Telescope Science Institute, which is operated by AURA for NASA; {\tt sbpy}, an {\tt astropy} affiliated package for small-body planetary astronomy \citep{mommert2019_sbpy};
a {\tt python} implementation of the \citet{cowan1979_cometsublimation} sublimation model \citep{vanselous21_ice_e20745b}; and
{\tt uncertainties} (version 3.0.2), a {\tt python} package for calculations with uncertainties by E.~O.\ Lebigot\footnote{\url{http://pythonhosted.org/uncertainties/}}.

\begin{acknowledgments}

We thank two anonymous reviewers for helpful feedback that improved this manuscript.
This work benefited from support from the International Space Science Institute, Bern, Switzerland, through the hosting and provision of financial support for an international team to discuss the science of main-belt comets.  This work is based on observations made with the NASA/ESA/CSA James Webb Space Telescope.  The data were obtained from the Mikulski Archive for Space Telescopes at the Space Telescope Science Institute, which is operated by the Association of Universities for Research in Astronomy, Inc., under NASA contract NAS 5-03127 for \jwst{}.  These observations are associated with \jwst{} General Observer Program 4250.
Support for this work was provided to H.H.H., J.W.N., M.S.P.K., and D.B.\ by NASA through grant NAS 5-03127 from the Space Telescope Science Institute, which is operated by the Association of Universities for Research in Astronomy, Inc., under contract NAS 5-26555.

H.H.H., M.S.P.K., J.P., S.S.S., and A.T.\ also acknowledge support from the NASA Solar System Observations program (Grant 80NSSC19K0869).  The work of J.P.\ was conducted at the Jet Propulsion Laboratory, California Institute of Technology, under a contract with the National Aeronautics and Space Administration (80NM0018D0004).
C.O.C. acknowledges support from the NASA CSSFP (grant No. 80NSSC26K0380), and Arthur and Jeanie Chandler. LINCC Frameworks is supported by Schmidt Sciences. C.O.C. also acknowledges support from the DiRAC Institute in the Department of Astronomy at the University of Washington. The DiRAC Institute is supported through generous gifts from the Charles and Lisa Simonyi Fund for Arts and Sciences.

The authors thank M.\ M.\ Knight for assistance in obtaining NTT observations. J.W.N acknowledges Benjamin N.L. Sharkey and Anicia Arredondo for useful discussions regarding reflectance spectroscopy. 

This work is based on observations obtained at the international Gemini Observatory (under Programs GS-2023A-LP-104, GS-2024A-Q-111, GN-2024B-Q-114, and GS-2024B-Q-113), a program of NSF NOIRLab, which is managed by the Association of Universities for Research in Astronomy (AURA) under a cooperative agreement with the U.S. National Science Foundation on behalf of the Gemini Observatory partnership: the U.S. National Science Foundation (United States), National Research Council (Canada), Agencia Nacional de Investigaci\'{o}n y Desarrollo (Chile), Ministerio de Ciencia, Tecnolog\'{i}a e Innovaci\'{o}n (Argentina), Minist\'{e}rio da Ci\^{e}ncia, Tecnologia, Inova\c{c}\~{o}es e Comunica\c{c}\~{o}es (Brazil), and Korea Astronomy and Space Science Institute (Republic of Korea).
Portions of this work were specifically enabled by observations made from the Gemini North telescope, located within the Maunakea Science Reserve and adjacent to the summit of Maunakea. We are grateful for the privilege of observing the Universe from a place that is unique in both its astronomical quality and its cultural significance.

This work is also based on observations obtained at the Hale Telescope at Palomar Observatory as part of a continuing collaboration between the California Institute of Technology, NASA/JPL, Yale University, and the National Astronomical Observatories of China.

This work is also based on observations obtained at the Lowell Discovery Telescope at Lowell Observatory.
Lowell is a private, non-profit institution dedicated to astrophysical research and public appreciation of astronomy and operates the LDT in partnership with Boston University, the University of Maryland, the University of Toledo, Northern Arizona University and Yale University.
The Large Monolithic Imager was built by Lowell Observatory using funds provided by the National Science Foundation (AST-1005313). The University of Maryland LDT observing team consists of Q.\ Ye, J.\ Bauer, A.\ Gicquel-Brodtke, T.\ Farnham, L.\ Farnham, C.\ Holt, M.\ S.\ P.\ Kelley, J.\ Kloos, and J.\ Sunshine.

This work is also based on observations collected at the European Organisation for Astronomical Research in the Southern Hemisphere under ESO programme 113.26J9.002. The authors thank telescope operator Monica Castillo for assistance in observation collection. 

The authors thank 
T.\ Barlow, K.\ Koviak, P.\ Nied, and other Palomar Observatory staff; 
NTT observatory staff; and
J.\ Andrews, R.\ Angeloni, P.\ Candia, R.\ Carrasco, J.\ Chavez, K.\ Chiboucas, A.\ Cikota, B.\ Cooper, E.\ Deibert, V.\ Firpo, J.\ Font-Serra, J.\ Fuentes, J.-E.\ Heo, S.\ Leggett, A.\ Lopez, L.\ Magill, C.\ Mart\'inez-V\'azquez, D.\ May, B.\ Miller, T.\ Mo{\v c}nik, S.\ Panda, P.\ Prado, M.\ Rawlings, R.\ Ruiz, T.\ Seccull, K.\ Silva, A.\ Smith, A.\ Stephens, S.\ Stewart, J.\ Thomas-Osip
and other Gemini Observatory staff
for their assistance in obtaining observations.

The DECam Legacy Surveys consist of three individual and complementary projects: the Dark Energy Camera Legacy Survey (DECaLS; Prop.\ ID 2014B-0404; PIs: D.\ Schlegel and A.\ Dey), the Beijing-Arizona Sky Survey (BASS; NOAO Prop.\ ID 2015A-0801; PIs: Z.\ Xu and X.\ Fan), and the Mayall z-band Legacy Survey (MzLS; Prop.\ ID 2016A-0453; PI: A.\ Dey). DECaLS, BASS and MzLS together include data obtained, respectively, at the Blanco telescope, Cerro Tololo Inter-American Observatory, NSF's NOIRLab; the Bok telescope, Steward Observatory, University of Arizona; and the Mayall telescope, Kitt Peak National Observatory, NOIRLab. Pipeline processing and analyses of the data were supported by NOIRLab and the Lawrence Berkeley National Laboratory (LBNL). The Legacy Surveys project is honored to be permitted to conduct astronomical research on Iolkam Du'ag (Kitt Peak), a mountain with particular significance to the Tohono O'odham Nation.

NOIRLab is operated by the Association of Universities for Research in Astronomy (AURA) under a cooperative agreement with the National Science Foundation. LBNL is managed by the Regents of the University of California under contract to the U.S. Department of Energy.

This project used data obtained with the Dark Energy Camera (DECam), which was constructed by the Dark Energy Survey (DES) collaboration. Funding for the DES Projects has been provided by the U.S.\ Department of Energy, the U.S.\ National Science Foundation, the Ministry of Science and Education of Spain, the Science and Technology Facilities Council of the United Kingdom, the Higher Education Funding Council for England, the National Center for Supercomputing Applications at the University of Illinois at Urbana-Champaign, the Kavli Institute of Cosmological Physics at the University of Chicago, Center for Cosmology and Astro-Particle Physics at the Ohio State University, the Mitchell Institute for Fundamental Physics and Astronomy at Texas A\&M University, Financiadora de Estudos e Projetos, Fundacao Carlos Chagas Filho de Amparo, Financiadora de Estudos e Projetos, Funda\c{c}$\tilde{a}$o Carlos Chagas Filho de Amparo {\`a} Pesquisa do Estado do Rio de Janeiro, Conselho Nacional de Desenvolvimento Cient{\'i}fico e Tecnol{\'o}gico and the Minist{\'e}rio da Ci{\^e}ncia, Tecnologia e Inova\c{c}$\tilde{a}$o, the Deutsche Forschungsgemeinschaft and the Collaborating Institutions in the Dark Energy Survey. The Collaborating Institutions are Argonne National Laboratory, the University of California at Santa Cruz, the University of Cambridge, Centro de Investigaciones En{\'e}rgeticas, Medioambientales y Tecnol{\'o}gicas-Madrid, the University of Chicago, University College London, the DES-Brazil Consortium, the University of Edinburgh, the Eidgen{\"o}ssische Technische Hochschule (ETH) Z{\"u}rich, Fermi National Accelerator Laboratory, the University of Illinois at Urbana-Champaign, the Institut de Ci{\`e}ncies de l'Espai (IEEC/CSIC), the Institut de F{\'i}sica d'Altes Energies, Lawrence Berkeley National Laboratory, the Ludwig Maximilians Universit{\"a}t M{\"u}nchen and the associated Excellence Cluster Universe, the University of Michigan, NSF's NOIRLab, the University of Nottingham, the Ohio State University, the University of Pennsylvania, the University of Portsmouth, SLAC National Accelerator Laboratory, Stanford University, the University of Sussex, and Texas A\&M University.

BASS is a key project of the Telescope Access Program (TAP), which has been funded by the National Astronomical Observatories of China, the Chinese Academy of Sciences (the Strategic Priority Research Program ``The Emergence of Cosmological Structures'' Grant XDB09000000), and the Special Fund for Astronomy from the Ministry of Finance. The BASS is also supported by the External Cooperation Program of Chinese Academy of Sciences (Grant 114A11KYSB20160057), and Chinese National Natural Science Foundation (Grant 12120101003, 11433005).

The Legacy Survey team makes use of data products from the Near-Earth Object Wide-field Infrared Survey Explorer (NEOWISE), which is a project of the Jet Propulsion Laboratory/California Institute of Technology. NEOWISE is funded by the National Aeronautics and Space Administration.

The Legacy Surveys imaging of the DESI footprint is supported by the Director, Office of Science, Office of High Energy Physics of the U.S.\ Department of Energy under Contract No.\ DE-AC02-05CH1123, by the National Energy Research Scientific Computing Center, a DOE Office of Science User Facility under the same contract; and by the U.S.\ National Science Foundation, Division of Astronomical Sciences under Contract No.\ AST-0950945 to NOAO.

We extend our gratitude to the two anonymous reviewers who provided the excellent feedback that improved our manuscript. 

\end{acknowledgments}

\renewcommand{\thesubsection}{\Alph{subsection}}

\appendix
\setcounter{figure}{0}
\renewcommand{\thefigure}{A\arabic{figure}}

\section{Composite Images from Ground-based Observations\label{section:appendix_groundbased_images}}

\begin{figure*}[ht]
    \centering
    \includegraphics[width=0.8\linewidth]{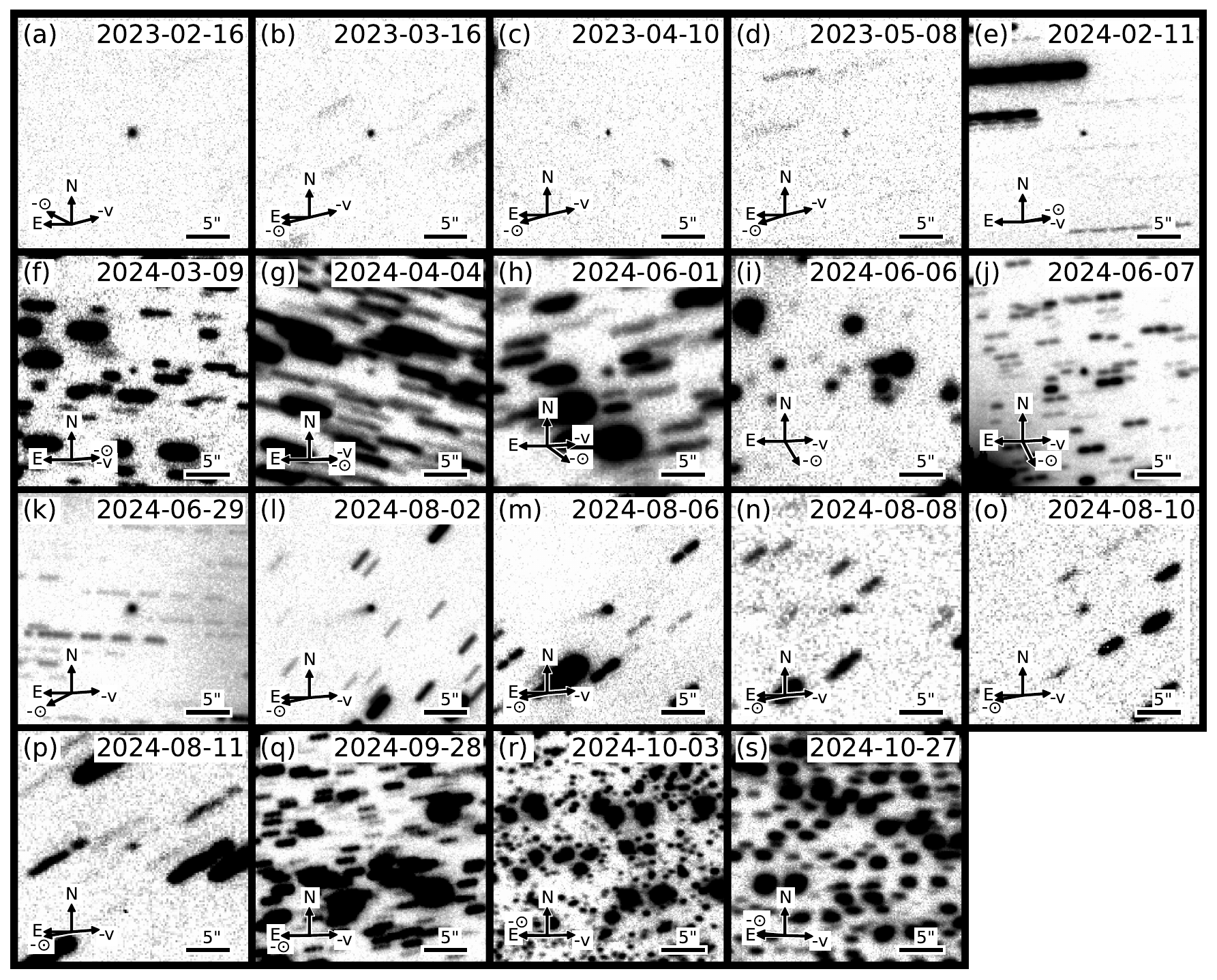}
    \caption{Composite images constructed from ground-based optical observations of 457P listed in Table~\ref{table:ground_observations_457p}.  Each panel includes the date of observation, a $5''$ scale bar to indicate the size of each image, and arrows denoting the directions of North (N), East (E), the antisolar vector projected on the sky ($-\odot$), and the negative heliocentric velocity vector projected on the sky ($-v$).
    }
    \label{fig:457p_optical_images}
\end{figure*}

\clearpage

\bibliography{merged_bib_clean,additonal_refs}
\bibliographystyle{aasjournalv7}

\end{CJK*}
\end{document}